\documentclass[%
 reprint,
 amsmath,amssymb,
 aps,
]{revtex4-2}

\usepackage{graphicx}
\usepackage{dcolumn}
\usepackage{bm}

\begin{document}


\title{Evidence of Spin-Glass state in Molecular Exchange-Bias System}%
\author{Suman Mundlia}
\author{Karthik V. Raman}%
 \email{kvraman@tifrh.res.in}
\affiliation{Tata institute of fundamental research, Hyderabad 500107, India
}





\begin{abstract}
 In conventional exchange-bias system comprising of a bilayer film of ferromagnet (FM) and antiferromagnet (AFM), investigating the role of spin-disorder and spin-frustration inside the AFM and at the interface has been crucial in understanding the fundamental mechanism controlling the exchange-bias -- an effect that leads to a horizontal shift in the magnetization hysteresis response of the FM. Similarly, in the recently reported monolayer molecular exchange-bias effect requiring no AFM layer, probing magnetic-disorder at the FM/molecule interface or inside the FM layer can provide new insights into the origin of molecular exchange-bias and the associated physics. In this article, by cooling the Fe/metal-phthalocyanine devices in oscillating magnetic field, we demonstrate a characteristic temperature dependent response of exchange-bias shift and ferromagnet coercivity that is supportive of a spin-glass behavior. Here, the origin of spin-glass is attributed to the spin-frustration created in the magnetic structure of the Fe layer, which was absent in our reference-Fe studies. These results highlight the strong influence of FM/molecule interface $\pi - d$ hybridization on the magnetic exchange interactions extending deeper into the FM layer.
 
\end{abstract}
\maketitle
\section{Introduction} 
Molecular exchange-bias effect \cite{gruber2015exchange,boukari2018disentangling,jo2018molecular,jo2020emergence,PhysRevApplied.14.024095} is a particular manifestation of ferromagnet/molecule spin-interface study \cite{sanvito2010molecular} in a widely investigated topic of interface-assisted molecular spintronics \cite{raman2014interface,atodiresei2014interface}. Here, complex charge transfer and hybridization effects between the molecular orbitals and the spin-polarised bands of the ferromagnet (FM) surface create new hybridized interface states that provide novel electronic and magnetic properties  \cite{Atodiresei2013ChemicalAV,raman2014interface, atodiresei2014interface}. Importantly, studies indicate the possibility to tune the magnetic-exchange interaction \cite{PhysRevLett.111.106805} and the magnetic anisotropy at the surface \cite{raman2014interface} with the latter receiving wider attention through the experimental reports of enhancement in film coercivity (H$_c$) \cite{raman2013interface,PhysRevB.87.041403,PhysRevLett.114.247203}. In recent studies, such a surface magnetic hardening response is found to be directly responsible for molecular exchange-spring \cite{PhysRevB.101.060408} and exchange-bias effect \cite{gruber2015exchange, boukari2018disentangling, jo2018molecular, jo2020emergence}. In the case of latter, the effect was recently shown to appear in a monolayer molecular film on FM \cite{PhysRevApplied.14.024095}, suggesting its mechanism to arise from the magnetization response of a hard-FM (surface)/soft-FM (bulk) bilayer system (see figure 1a) which is evidently different from the previous studies of exchange-bias effect in FM/AFM bilayers. However, despite these differences, existing literature for FM/AFM systems \cite{nogues1999exchange,zhang2016epitaxial} can provide a better understanding of the possible magnetization dynamics that may be playing a role in molecular exchange-bias effect.

\par In conventional FM/AFM systems \cite{PhysRev.105.904}, a number of parameters are known to influence exchange bias -- these include magnetic anisotropy, crystal structure \cite{yuan2016crystal}, magnetic domains \cite{PhysRevB.68.214420,nowak2001domain,PhysRevB.59.3722}, interface roughness \cite{PhysRevB.94.104425, MAITRE2012403,PhysRevB.35.3679}, spin-configuration \cite{PhysRevB.63.174422, PhysRevLett.79.1130} and exchange coupling \cite{fan2013exchange,zhou2017robust,ali2007exchange}. However, in molecular exchange-bias effect, absence of an AFM layer allow us to reduce the above parameter space, thereby limiting our study to the properties of interface and the underneath FM layer. In particular, surface roughness, magnetic domain state, hybridization induced anisotropy and magnetic exchange coupling near the surface are expected to play a dominant role in molecular exchange-bias. 
\begin{figure}[t]
    \centering
\includegraphics[width=0.48\textwidth]{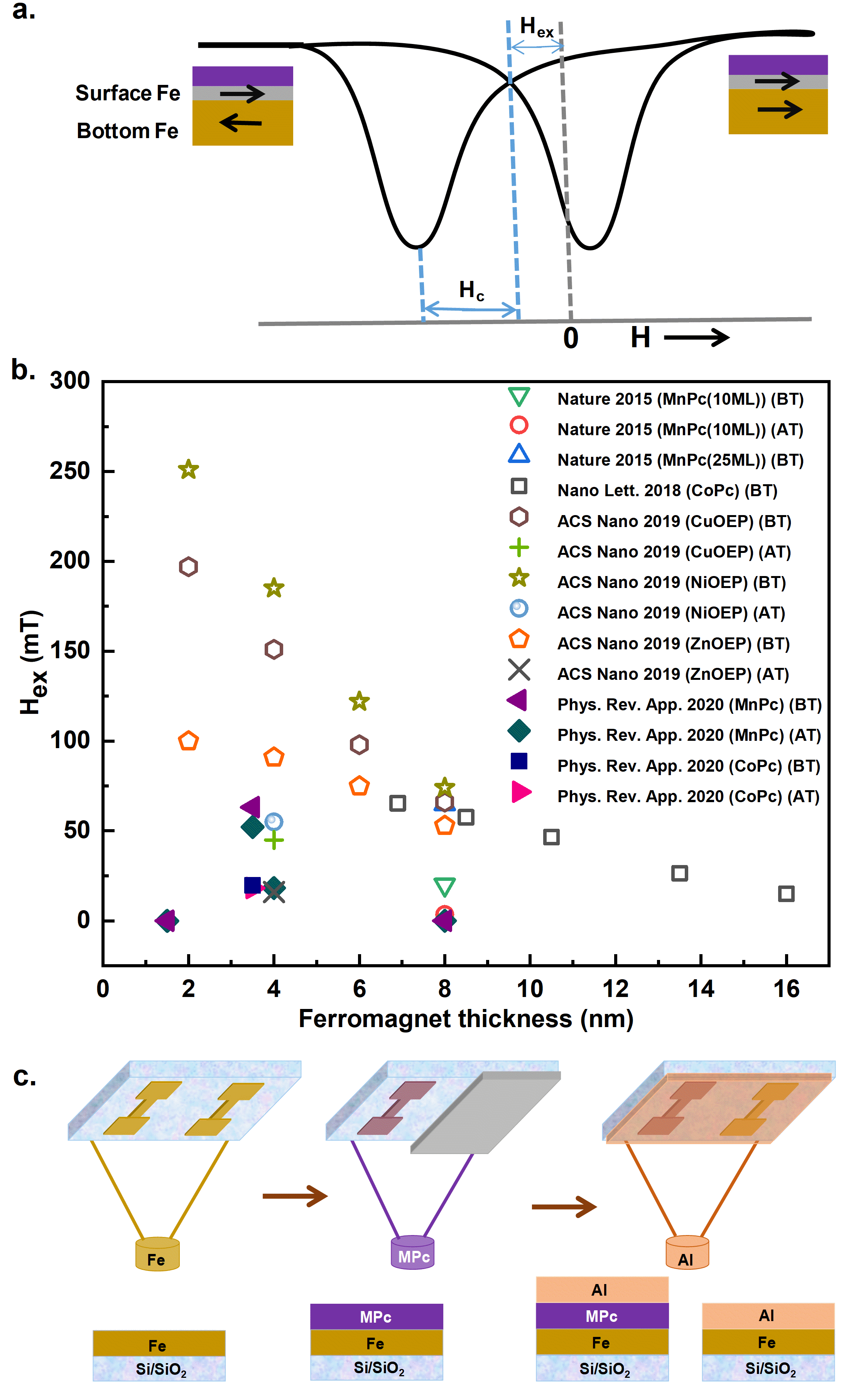} 
\caption{(a) Schematic of molecular exchange bias in a hard-surface/soft-Fe bilayer showing magnetization switching of bottom Fe (yellow) relative to the pinned surface Fe (grey) causing a horizontal shift in AMR by H$_{ex}$.
(b) Figure shows the literature data of molecular exchange bias showing H$_{ex}$ vs FM layer thickness measured before training (BT) and after training (AT) cycles. (c) Schematic showing deposition procedure of reference-Fe and  Fe/MPc devices in a single deposition run using shadow mask.}
   \label{fig:1}
\end{figure} 
\par In literature, molecular exchange-bias effect is reported in different thickness range of FMs and also of the molecule \cite{gruber2015exchange, boukari2018disentangling, jo2018molecular,jo2020emergence,PhysRevApplied.14.024095}. Here, thicker molecular spin-chains (acting as the AFM layer) are believed to couple with the surface hard layer during the field cooling process and temporarily stabilize the surface magnetism \cite{gruber2015exchange,boukari2018disentangling,jo2018molecular,jo2020emergence}. However, during magnetization hysteresis studies, the stability of the molecular spin-chain alignment is weakened (or lost), leading to a strong dependence of exchange-bias shift (H$_{ex}$) with training cycle. This behavior is noted in figure 1b for a number of previous published reports on molecular exchange-bias showing the value of H$_{ex}$ during 1$^{st}$ cycle with a significant drop after sufficient training period (AT). In thicker films ($>$ 4 nm), exchange bias is weak with strong training effect leading to an almost complete loss of H$_{ex}$. However, recent studies in thinner FM films ($<$ 4 nm) show sufficient enhancement in H$_{ex}$ and robustness in their training response \cite{jo2018molecular,jo2020emergence}. More importantly, as shown in figure 1b (solid symbols), monolayer molecular study \cite{PhysRevApplied.14.024095} provides considerably weaker training effect perhaps due to the absence of such molecular spin-chains. In conventional FM/AFM systems, similar enhancement in H$_{ex}$ at lower temperatures and in thinner AFM layers ($<$ 2 nm) is observed  \cite{PhysRevB.68.214420}. Here, they are attributed to the formation of perpendicular domain walls (perpendicular to the surface) in the AFM layer that become energetically unfavorable upon increasing the AFM layer thickness \cite{PhysRevB.68.214420, PhysRevB.35.3679}. Lack of an AFM layer in molecular exchange bias may rule out the above mechanism. However, similar possibility of perpendicular domain wall formation at the surface of FM that freeze with cool down and contribute to enhancement in H$_{ex}$ of an FM/molecule system may not be completely ruled out. Especially in the case of thinner FM films ($< \sim$ 4 nm), we expect these domain walls to extend deeper into the bottom FM layer. Occurrence of such perpendicular domain walls will in-principle cause H$_{ex}$ to depend on the strength of in-plane magnetic field cooling since such fields can modulate the density of perpendicular domain walls. Up till now, this was however not observed in our previous studies \cite{PhysRevApplied.14.024095} indicating that perpendicular domain walls, although likely to be present, do not play a limiting role in magnetization dynamics of our devices. In effect, the phenomena of surface magnetic hardening naturally favor the formation of planar domain walls, parallel to the film surface, during the magnetization switching of the bottom Fe layer \cite{mauri1987simple}. This switching response can be modeled using the Stoner-Wolfarth (SW) model \cite{stoner1948mechanism}, as was done in Ref. \cite{PhysRevApplied.14.024095}, that provided a qualitative understanding of the role of surface anisotropy in the origin of H$_{ex}$. Nevertheless, it is possible that both domain walls coexist in thinner films \cite{wee2001temperature,PhysRevB.66.014430} which may require sensitive spectroscopy techniques for careful investigation. The fact that domain wall dynamics do play a contributory role in molecular exchange-bias is evident from the study of even thinner films of Fe ($<$ 2 nm) showing no H$_{ex}$ but only an enhancement in H$_c$ \cite{PhysRevApplied.14.024095}. This response can be explained by the unfavorable energy conditions for the formation of domain walls. 
\par In this article, we focus our study towards the investigation of magnetization response and possible domain structure dynamics in Fe/MnPc devices after cooling them in oscillating magnetic fields. Such a procedure gives us access to magnetic disorder studies showing spin-glass behavior \cite{schlenker1986magnetic,mydosh1993spin} that was not possible in earlier studies of zero-field or field-cooled measurements. Our studies reveal the presence of surface $\pi - d$ hybridization induced strong modification in magnetic exchange coupling in the FM layer that may provide additional insights into the recent report of molecular crane-pulley effect \cite{PhysRevApplied.14.024095}.
\begin{figure*}[t]
    \centering
\includegraphics[width=0.98\textwidth]{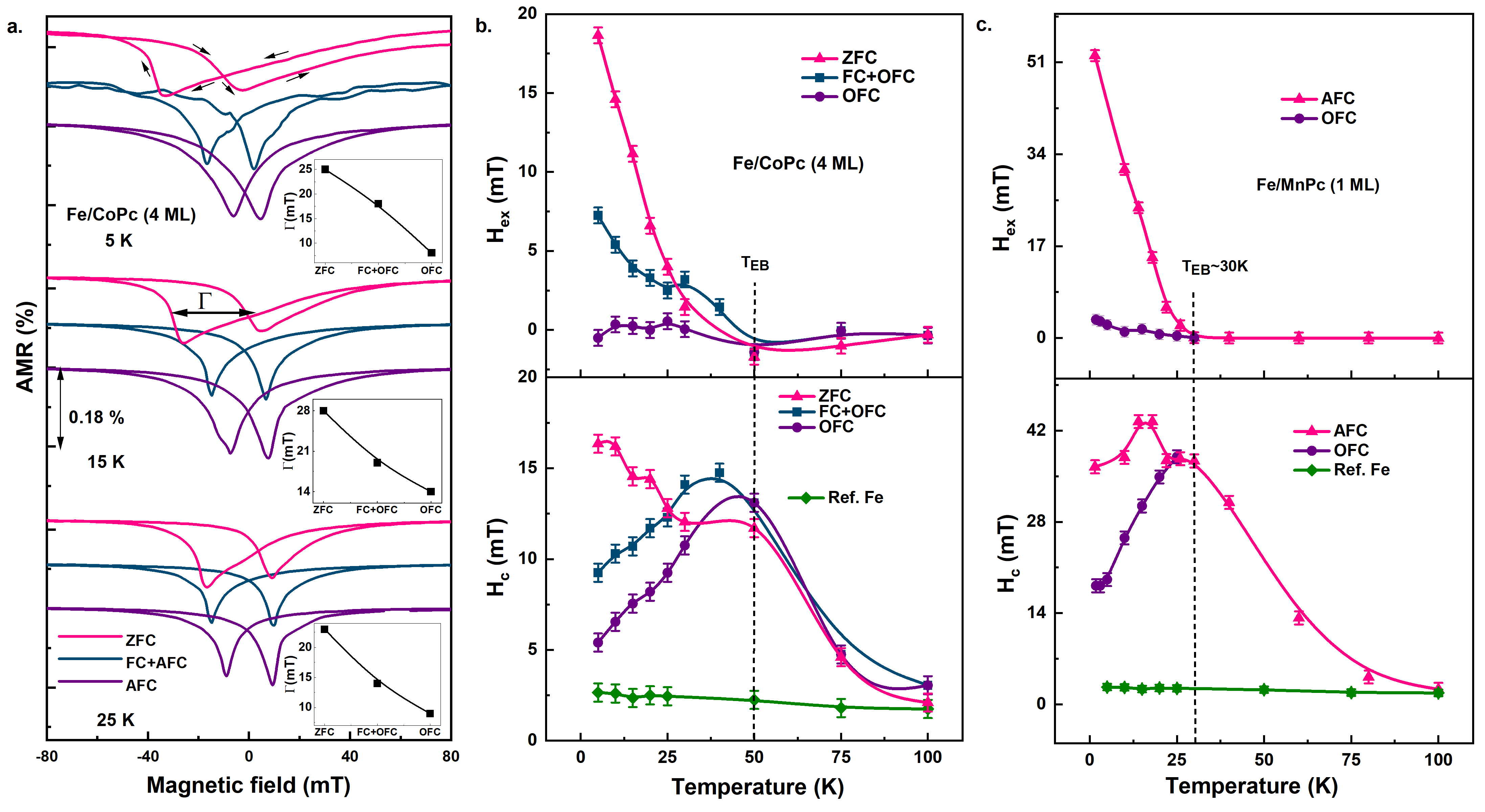} 
\caption{(a) AMR in \% of Fe (3.5 nm)/CoPc (4 ML) devices at 5 K (top), 15 K (middle), and 25 K (bottom) in ZFC (pink), FC+AFC (green) and AFC (purple) conditions. In the inset of each AMR plot magnitude of $\Gamma$ (shown in the middle) is plotted for different field cooling.  (b) H$_{ex}$ vs T (top) and  H$_{c}$ vs T (bottom) of Fe (3.5 nm)/CoPc (4 ML) device extracted from the AMR plots in ZFC (pink triangle), FC + OFC (green squares) and OFC (purple circles). (c) H$_{ex}$ vs T (top) and H$_{c}$ vs T (bottom) of Fe (3.5 nm)/MnPc (1 ML) device in ZFC (pink triangle) and OFC (purple circles).}
   \label{fig:2}
\end{figure*} 
\section{Experimental details}
\par For our current study, two kinds of samples, \emph{viz.} reference-Fe and Fe/metal-phthalocyanine (MPc, M: Co and Mn) \cite{lu2016optically} bilayers were prepared in an ultra-high vacuum thin film deposition system. In figure 1c, we show the deposition procedure of our patterned films using shadow mask technique. First, the Fe film is grown simultaneously on the two substrates. One of the substrate is then covered by an in-situ mask before the molecule is evaporated. Finally, both the films are capped with Al (capping layer) to prevent oxidation of underneath layers. Analysis of a complete capping layer coverage is performed using Energy dispersive X-ray elemental mapping. These details can be found in Ref. \cite{PhysRevApplied.14.024095}. Such a growth procedure allow us to avoid any variation in the Fe film thickness between the reference-Fe and Fe/MPc devices that are loaded together in our variable temperature insert cryostat. Here, the samples are cooled to the base temperature of 2 K under different field cooling conditions and isothermal anisotropic magnetoresistance (AMR) measurements are recorded during the warm-up.
\section{Results and Discussion}
\par In our previous studies \cite{PhysRevApplied.14.024095}, we observed that zero-field cooling (ZFC) or field-cooling (FC) the Fe/MPc devices in different magnetic fields (ranging from 0 T to 0.6 T) do not lead to any visible change in H$_{ex}$ suggesting the dominant role of planar domain walls in magnetization switching dynamics. In comparison, reference-Fe films do not show any H$_{ex}$ even under FC conditions with a significantly lower H$_c$. In the present work, we have extended the AMR studies to oscillating field cooling (OFC) conditions --wherein the devices are cooled in an in-plane oscillating field with an amplitude of 50 mT at $\sim$0.1 Hertz. Figure 2a shows the temperature dependence of AMR in Fe (3.5 nm)/CoPc (4ML) devices acquired during warm-up in ZFC, OFC and FC + OFC (i.e. the field was made to oscillate between 70 mT and -30 mT) cooling conditions. Clearly, these cooling conditions lead to a change in the value of H$_{ex}$ dropping down to nearly zero for OFC measurements (see figure 2b). Additionally, the magnitude of H$_c$ and the width ($\Gamma$) of the AMR drops (inset of figure 2) at H$_c$ are found to decrease in moving from ZFC to OFC plots. In contrast, the AMR plots of our reference-Fe films after OFC show similar value of H$_c$ with a marginal increase in $\Gamma$ compared to the ZFC measurements (not shown). The above behavior in AMR response of Fe/MPc devices suggest that OFC conditions lead to changes in the magnetic structure that could arise in the top hard-surface layer and possibly extend into the bottom Fe layer. 

\par We now examine the possible effects of OFC on the domain dynamics-- this may include (a) random orientation of the hard-surface layer domains mostly parallel or anti-parallel to the magnetic field axis leading to an overall drop in H$_{ex}$ and/or (b) drop in average domain size in the hard-surface layer and/or (c) changes in the magnetic domain structure of the bottom soft-Fe layer. If we assume that OFC only leads to case (a) with no change in domain size dynamics, we would expect a similar value of H$_c$ and an increase in the width $\Gamma$ of the AMR peaks compared to ZFC measurements (as noted in reference-Fe films). This is also suggested by our simulation studies using the SW model (see Appendix B). However, our experiments show a drop in H$_c$ and $\Gamma$ suggesting domain size dynamics to possibly contribute. In order to investigate the above in greater detail, in figure 2b \& 2c, we plot the temperature dependence of the extracted values of H$_{ex}$ and H$_c$ for the Fe/CoPc and Fe/MnPc device cooled under different conditions. Here, the enhancement in H$_c$, above the exchange-bias onset temperature (T$_{EB}$), relative to reference-Fe is attributed to surface magnetic hardening effects giving rise to a molecular crane-pulley response below T$_{EB}$ \cite{PhysRevApplied.14.024095}. In figure 2, in the case of ZFC measurements, H$_c$ monotonically increases with decreasing temperature with an increase in H$_{ex}$ below T$_{EB}$ of $\sim$50 K in CoPc devices and 30 K in the case of MnPc devices. Here, based on our understanding of FM/AFM \cite{nayak2020exchange,PhysRevB.68.214420,ding2013interfacial} or FM/spin-glass exchange-bias system \cite{ali2007exchange}, the enhancement in H$_c$ relative to the reference-Fe can be understood to arise from two additive but competing effects -- surface exchange induced magnetic anisotropy (K$^i_{ex}$) in the bottom Fe layer and softer domains or frustrated spin centers in the hard surface layer matrix that couple with the bottom Fe layer and partially rotate upon magnetization reversal. Generally, the latter effect depends on interface-disorder and spin-frustration and its strong presence leads to a drop in H$_c$ at lower temperatures caused by surface domain freezing, whereby they stop contributing to the H$_c$ enhancement \cite{ali2007exchange,PhysRevB.68.214420,xiantiferro2000}. While on the other hand, K$^i_{ex}$ enhances at lower temperature contributing to an increase in H$_{c}$, as noted in previous studies \cite{PhysRevB.63.174422} and in our ZFC data.

\par In the case of FC + OFC procedure (see figure 2b), H$_{ex}$ shows a decrease during warm-up with a plateau-like response at around 30 K. Additionally, at 2K, its value remain lower than in the case of ZFC. Interestingly, in this temperature range, H$_c$ shows an opposite trend with warm-up compared to the ZFC data. Here, H$_c$ at 2K is at $\sim$ 8 mT that is considerably lower than the ZFC value (16 mT), and show an increase with warm-up reaching $\sim$ 14 mT. In the 30K - 50K temperature range, FC + OFC show a larger value of H$_c$ than in ZFC hinting towards the possible presence of softer domains frustrated in the magnetically hard surface matrix that couple to the bottom Fe layer during magnetization reversal thereby enhancing its H$_c$. At lower temperatures, these domains and the associated domain walls may remain frozen and do not (or weakly) contribute to H$_{ex}$.  Such a trend in H$_c$ and H$_{ex}$ may suggest a stronger interfacial spin-glass contribution to the magnetization switching dynamics as typically observed in conventional FM/spin-glass exchange-bias systems \cite{ali2007exchange}. Ideally, if this was the only scenario, we would expect OFC measurements (see figure 2a and 2b) to give a stronger enhancement in H$_c$, just below T$_{EB}$, over the values in ZFC and FC + OFC measurements. However, our results in figure 2b do not support such a response suggesting that the origin of non-monotonic temperature dependence of H$_c$ in OFC may not be completely driven by a conventional interfacial spin-glass behavior caused by the formation of softer (or superparamagnetic) smaller surface domains or isolated spin centers. 

\par On the contrary, in OFC, lack of H$_{ex}$ in the complete temperature range below T$_{EB}$ suggest that the spin-frustration caused by OFC is strong and must be creating a strongly pinned frustrated domain matrix at the surface, forming enhanced density of perpendicular domain walls that extend deeper into the bottom Fe layer. Formation of these frozen domains walls will, in-principle, also make domain wall motion harder during magnetization switching of the bottom Fe layer suggesting an enhancement in H$_c$ and $\Gamma$. This is however contrary to our experimental observations implying that the above domain dynamics also do not explain the temperature dependent AMR response in the devices. Ideally, we expect that OFC does not affect the local strength of surface hybridization induced K$_{UD}$ (in per unit interface area) since the energy scales involved here are very different with the hybridization effects being much stronger than the magnetic energy. We believe a better measure of average response from the unidirectional exchange anisotropy over the interface is not by extracting H$_{ex}$ which in this case is zero, but by probing the presence of such anisotropy using in-plane angle-dependent AMR studies. These studies allow us to analyse the asymmetric magnetization switching response with in-plane field rotation caused by the presence of interfacial unidirectional anisotropy (see appendix C). Interestingly, our experimental measurements show a difference in response compared to the conventional FM/AFM system i.e. Fe/FeO bilayer device. In our Fe/FeO system, the asymmetry in the AMR response disappear (appear) when the exchange-bias is absent (present) (see figure 5d). On the contrary, the Fe/MPc devices continue to show asymmetric AMR response even when the extracted H$_{ex} = 0$ implying the presence of interfacial unidirectional anisotropy of similar strength as in ZFC (see figure 5b). These observations, different from conventional FM/AFM systems, seem to suggest a different magnetic structure in Fe/MPc devices which may require further exploration in the future. 

\par The suggested existence of a strong interfacial exchange anisotropy but with a drop in H$_c$ and $\Gamma$ of the bottom Fe layer in OFC studies may indicate a relative loss in the magnetic anisotropy of the bottom Fe layer. The fact that H$_c$ remain higher than the value in reference-Fe, even at 2 K, suggest that the effects of K$_{ex}^i$ has reduced but is not completely absent. Such a response is often observed in reentrant spin-glass phases in weak ferromagnets \cite{RevModPhys.58.801,mydosh1993spin,pradheesh2012,PhysRevB.64.184432}. Here, a low-temperature mixed state is formed in which the spin-glass behavior coexists along with the long-range ferromagnetic order. These phases form due to weakening of magnetic exchange strength leading to cluster freezing at lower temperatures contributing to the spin-glass response. In our Fe/MPc devices, formation of such a mixed state with a spin-glass nature that is decoupled from the exchange coupling of the top surface (\textit{i.e.} K$_{UD}$) corroborate well with the experimental observation of a reduction in H$_c$ and $\Gamma$ in OFC measurements with no H$_{ex}$. With an increase in temperature, cluster freezing weakens leading to a dominant ferromagnetic phase with an increase in magnetic anisotropy. This leads to an increase in H$_c$ with an overall no change in H$_{ex}$ since the domains in the frustrated surface matrix grow equally for the parallel and the anti-parallel magnetization alignments. Hence, these above results highlight the formation of an unexpected spin-glass magnetic layer due to spin-frustration introduced at the molecule-absorbed surface offering new interesting physics at the fundamental level. Importantly, the work demonstrates the possibility to achieve strong modification in the strength of magnetic exchange coupling in an FM layer that is tunable by field cooling procedures. We believe, in the future, more sensitive techniques such as depth-resolved neutron reflectometry, ac-susceptibility, and M\"{o}ssbauer spectroscopy \cite{PhysRevB.64.184432} studies may provide direct insights into the rich magnetization dynamics in such FM/molecule hybrid bilayers.
\section{Conclusion}
\par In conclusion, OFC studies in FM/molecule devices has supported the formation of an unexpected spin-glass magnetic structure in the FM layer that seem to dominate the temperature dependent response of magnetization dynamics. These observations provide new insights into the strong effect of surface $\pi-d$ hybridization on the strength (weakening) of magnetic exchange interactions extending deeper into the FM layer and bringing them closer to the critical threshold to induce a long-range magnetic order. We believe our present work provides new insights into the important role of magnetic exchange interactions at the surface that may help to understand the origin of molecular exchange bias in tailoring them for potential device applications in 2D spintronics.

\section{Appendix}
\subsection{Modeling magnetization switching of bottom Fe layer with exchange-bias}

We model the magnetization switching response of a hard-FM/soft-FM bilayer system by considering the top hard-FM layer to be pinned along its internal magnetization axis which is responsible for the exchange-bias coupling on the bottom soft-FM layer. Stoner-Wolfarth (SW) model \cite{stoner1948mechanism} is used to model the magnetization switching response of a domain in the bottom soft-FM layer. 
\begin{figure}[h]
    \centering
\includegraphics[width=0.25\textwidth]{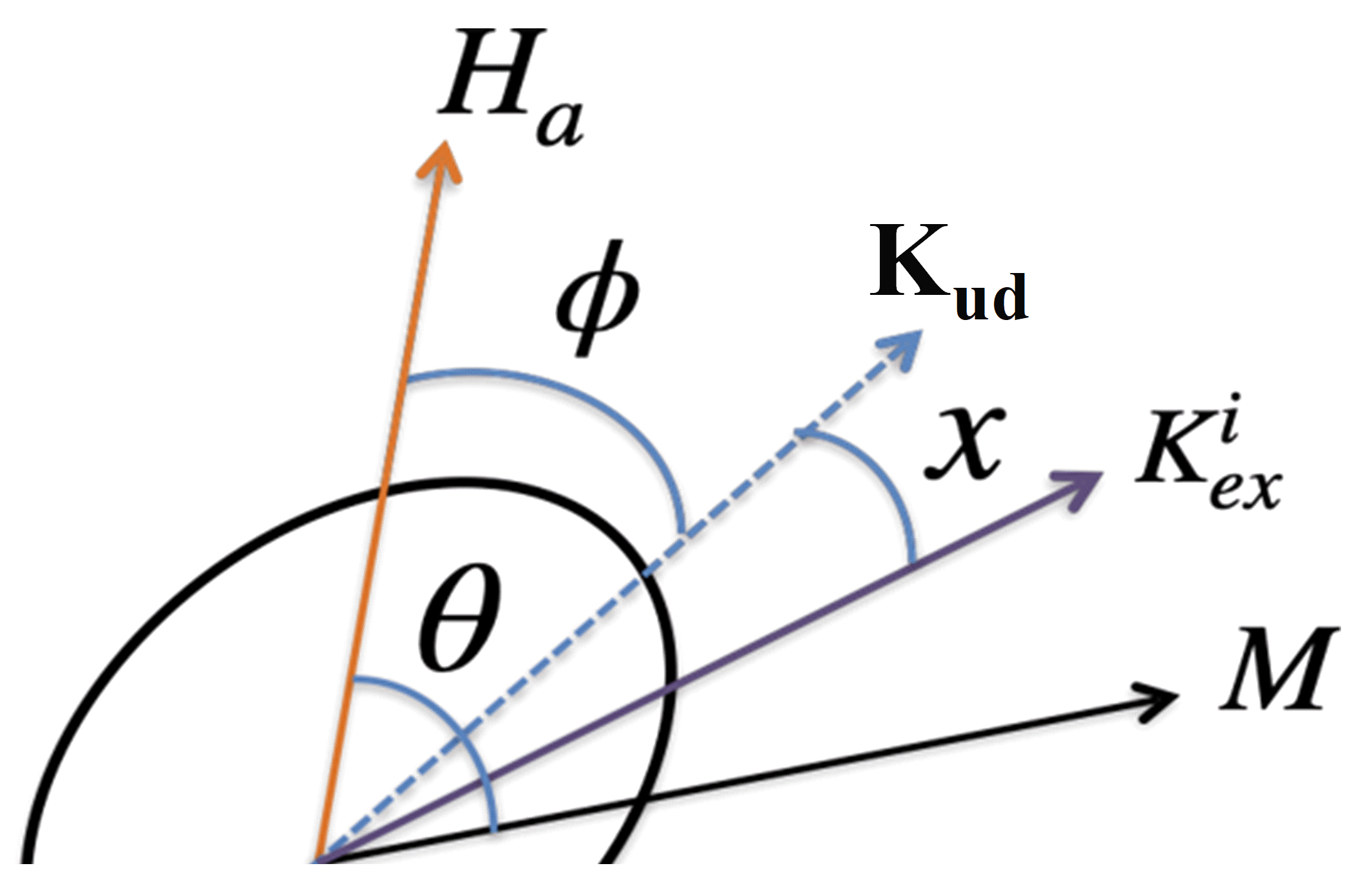} 
\caption{Schematic shows the representation of the SW model for simulating in-plane angular-dependent AMR.}
   \label{fig:3}
\end{figure} 
Here, we assume that the contribution of domain wall motion towards the value of H$_c$ is much lesser compared to the H$_c$ enhancement (relative to the value in reference-Fe) resulting from the surface hybridization induced magnetic anisotropy, K$^i_{ex}$. The interfacial exchange-bias coupling is introduced in the model by assuming an unidirectional exchange anisotropy K$_{UD}$ (see figure 3). K$_{ex}^i$ was previously shown to be biaxial in nature \cite{PhysRevApplied.14.024095}. In the following analysis, the results are independent of the nature of K$_{ex}^i$ but for convenience we consider a biaxial anisotropy term in the SW model. Here, the free energy per unit area (E) of bottom Fe layer can be estimated by \cite{nogues1999exchange}:
\begin{multline*}
E = -K_{UD}.cos(\theta-\phi) - K^i_{ex}.cos(4(\theta-\phi-x))\\
-\mu_oH_a.M_{Fe}.t_{Fe}.cos(\theta)
\end{multline*}
where $\theta$ is the angle between M and H$_a$, $x$ is the angle between K$_{UD}$ and K$_{ex}^i$ anisotropy axes (see figure 3). Here, the contribution of the internal anisotropy of the bottom Fe film is neglected and a random orientational absorption of the MPc molecules on Fe surface is considered to simulate the average AMR response, given by $<$AMR$>$ = $<cos^2\theta>$ - 1, where $< >$ represents an average over the in-plane azimuthal rotation of angle $x$.

\subsection{Effect of randomly oriented hard surface domains on the AMR response of bottom Fe layer}
\begin{figure}[t!]
    \centering
\includegraphics[width=0.45\textwidth]{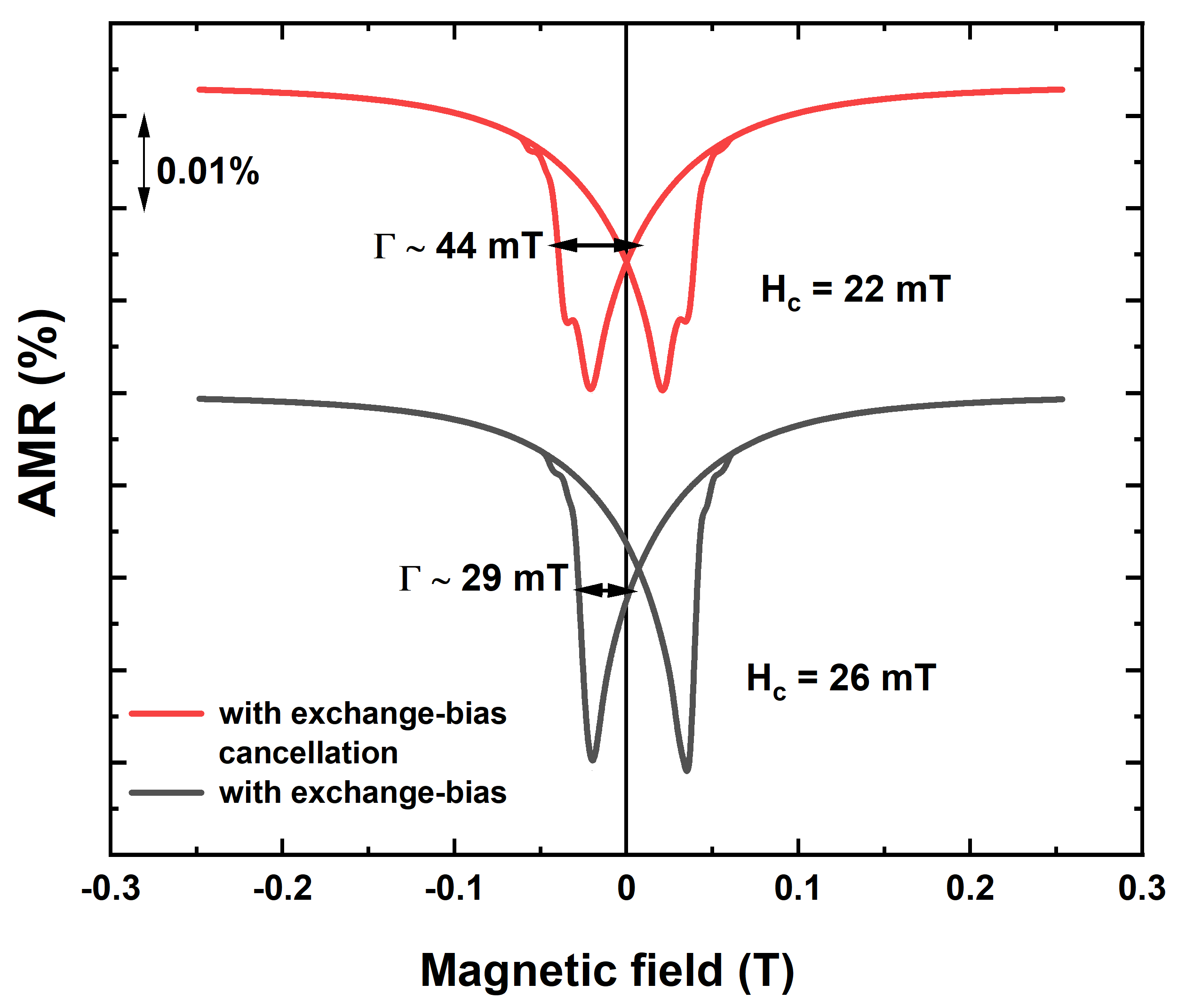} 
\caption{Simulated AMR plots for bottom Fe film using K$_{UD}$= 70 $\pm$ 10 mT showing exchange bias (bottom black) in uni-directional response of K$_{UD}$ in all the domains and a domain-averaged response over parallel and anti-parallel direction of K$_{UD}$ in different domains (top red) i.e. average over K$_{UD}$ = 70 and -70 mT domains.}
\label{fig:4}
\end{figure} 
In order to model the effect of OFC on the AMR response, we first consider the scenario that OFC only leads to random orientation of the surface domains primarily along the parallel or anti-parallel direction to the magnetic field axis. Here, we consider that the magnitude of K$_{UD}$ in each domain is unaffected by OFC and its strength is determined primarily by the $\pi-d$ hybridization. For such a scenario, we plot the simulated AMR response in figure 4 which shows similar value of H$_c$ and an enhancement in $\Gamma$ with no H$_{ex}$ in comparison to the case when the domains are aligned only in the parallel direction. 
\subsection{In-plane angle-dependent AMR studies}

\begin{figure}[h!]
    \centering
\includegraphics[width=0.48\textwidth]{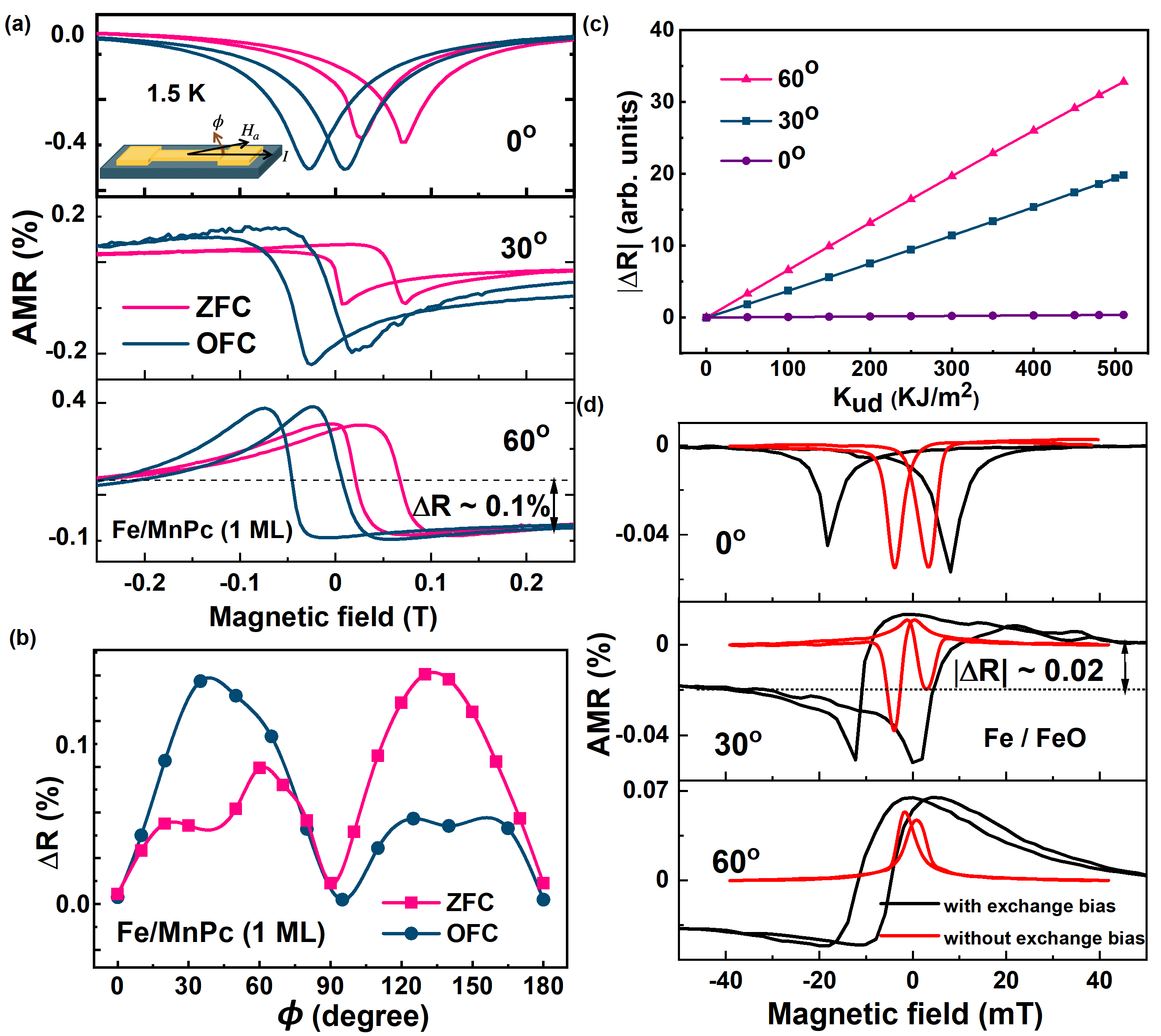} 
\caption{ (a) Angle ($\phi$ values in inset) dependent AMR plots of Fe (3.5 nm)/MnPc (1 ML) device in ZFC (pink) and OFC (green) conditions. Top-panel inset shows the device measurement geometry. (b) Magnitude of $\lvert \Delta R \rvert$ (shown in (a)) vs $\phi$ in ZFC (pink) and OFC (green) showing similar strength. (c) $\lvert \Delta R \rvert$ vs K$_{ud}$ extracted using the SW model is plotted for different values of $\phi$: 0$^o$ (purple circles), 30$^o$ (green squares), and 60$^o$ (pink triangles). (d) Angle ($\phi$ values in inset) dependent AMR plots of Fe/FeO device show $\lvert \Delta R \rvert$ only in the presence of exchange bias.}
\label{fig:5}
\end{figure} 
\par Angle-dependent AMR studies of FM/AFM devices in presence of exchange-bias (or K$_{UD}$) often given rise to asymmetric features due to the difference in magnetization rotation pathways in the ascending and descending branches of field sweep \cite{fitzsimmons2000relation,krivorotov2002relation}. When imposed over the internal anisotropy of the FM layer, the asymmetric features can take different trend with rotation of applied magnetic field. Figure 5a shows the angle-dependent AMR measurements performed on Fe/MnPc devices after cooling them in ZFC and OFC conditions. Here, in the case of OFC measurements, no H$_{ex}$ is observed. In our analysis, the asymmetric response with $\phi$ is captured by $\lvert \Delta R \rvert$ (shown in figure 5b) whose origin can be established directly with K$_{UD}$. This is shown in figure 5c where we plot the value of $\lvert \Delta R \rvert$ extracted from the AMR plots simulated at different $\phi$s by varying K$_{UD}$. Clearly, for all values of $\phi$, $\lvert \Delta R \rvert$ approaches zero with a reduction in K$_{UD}$ suggesting its origin to the unidirectional interface exchange-anisotropy. In addition, we also experimentally confirm the dependency of  $\lvert \Delta R \rvert$ on K$_{UD}$ by performing angle dependent AMR studies in conventional FM/AFM (Fe/FeO) system (shown in figure 5d). Here, Fe (10nm)/FeO devices are prepared by in-situ oxygen plasma oxidation after Fe growth. In these devices, $\lvert \Delta R \rvert$ at $\phi \neq 0$ is observed only when H$_{ex}$ is non-zero.

\par Surprisingly, in our Fe/MnPc devices, we observe $\lvert \Delta R \rvert$ to exist even after OFC conditions (showing negligible H$_{ex}$) with its magnitude similar to the values observed after ZFC. These above observations seem to suggest that (i) AMR response in Fe/MPc devices is different from conventional FM/AFM bilayers and (ii) OFC procedure do not modify the local strength of interfacial unidirectional exchange anisotropy i.e. K$_{UD}$. We believe further investigations using neutron reflectometry techniques may be needed to examine these differences in greater details.
\section{Acknowledgements}
We acknowledge extra mural funding from the Department of Atomic Energy, TIFR Hyderabad and Science and Engineering Research Board of the Government of India (SERB sanction no. ECR/2015/000199). 


\begin{thebibliography}{44}%
\makeatletter
\providecommand \@ifxundefined [1]{%
 \@ifx{#1\undefined}
}%
\providecommand \@ifnum [1]{%
 \ifnum #1\expandafter \@firstoftwo
 \else \expandafter \@secondoftwo
 \fi
}%
\providecommand \@ifx [1]{%
 \ifx #1\expandafter \@firstoftwo
 \else \expandafter \@secondoftwo
 \fi
}%
\providecommand \natexlab [1]{#1}%
\providecommand \enquote  [1]{``#1''}%
\providecommand \bibnamefont  [1]{#1}%
\providecommand \bibfnamefont [1]{#1}%
\providecommand \citenamefont [1]{#1}%
\providecommand \href@noop [0]{\@secondoftwo}%
\providecommand \href [0]{\begingroup \@sanitize@url \@href}%
\providecommand \@href[1]{\@@startlink{#1}\@@href}%
\providecommand \@@href[1]{\endgroup#1\@@endlink}%
\providecommand \@sanitize@url [0]{\catcode `\\12\catcode `\$12\catcode
  `\&12\catcode `\#12\catcode `\^12\catcode `\_12\catcode `\%12\relax}%
\providecommand \@@startlink[1]{}%
\providecommand \@@endlink[0]{}%
\providecommand \url  [0]{\begingroup\@sanitize@url \@url }%
\providecommand \@url [1]{\endgroup\@href {#1}{\urlprefix }}%
\providecommand \urlprefix  [0]{URL }%
\providecommand \Eprint [0]{\href }%
\providecommand \doibase [0]{https://doi.org/}%
\providecommand \selectlanguage [0]{\@gobble}%
\providecommand \bibinfo  [0]{\@secondoftwo}%
\providecommand \bibfield  [0]{\@secondoftwo}%
\providecommand \translation [1]{[#1]}%
\providecommand \BibitemOpen [0]{}%
\providecommand \bibitemStop [0]{}%
\providecommand \bibitemNoStop [0]{.\EOS\space}%
\providecommand \EOS [0]{\spacefactor3000\relax}%
\providecommand \BibitemShut  [1]{\csname bibitem#1\endcsname}%
\let\auto@bib@innerbib\@empty
\bibitem [{\citenamefont {Gruber}\ \emph {et~al.}(2015)\citenamefont {Gruber},
  \citenamefont {Ibrahim}, \citenamefont {Boukari}, \citenamefont {Isshiki},
  \citenamefont {Joly}, \citenamefont {Peter}, \citenamefont {Studniarek},
  \citenamefont {Da~Costa}, \citenamefont {Jabbar}, \citenamefont {Davesne}
  \emph {et~al.}}]{gruber2015exchange}%
  \BibitemOpen
  \bibfield  {author} {\bibinfo {author} {\bibfnamefont {M.}~\bibnamefont
  {Gruber}}, \bibinfo {author} {\bibfnamefont {F.}~\bibnamefont {Ibrahim}},
  \bibinfo {author} {\bibfnamefont {S.}~\bibnamefont {Boukari}}, \bibinfo
  {author} {\bibfnamefont {H.}~\bibnamefont {Isshiki}}, \bibinfo {author}
  {\bibfnamefont {L.}~\bibnamefont {Joly}}, \bibinfo {author} {\bibfnamefont
  {M.}~\bibnamefont {Peter}}, \bibinfo {author} {\bibfnamefont
  {M.}~\bibnamefont {Studniarek}}, \bibinfo {author} {\bibfnamefont
  {V.}~\bibnamefont {Da~Costa}}, \bibinfo {author} {\bibfnamefont
  {H.}~\bibnamefont {Jabbar}}, \bibinfo {author} {\bibfnamefont
  {V.}~\bibnamefont {Davesne}}, \emph {et~al.},\ }\bibfield  {title} {\bibinfo
  {title} {Exchange bias and room-temperature magnetic order in molecular
  layers},\ }\href@noop {} {\bibfield  {journal} {\bibinfo  {journal} {Nature
  materials}\ }\textbf {\bibinfo {volume} {14}},\ \bibinfo {pages} {981}
  (\bibinfo {year} {2015})}\BibitemShut {NoStop}%
\bibitem [{\citenamefont {Boukari}\ \emph {et~al.}(2018)\citenamefont
  {Boukari}, \citenamefont {Jabbar}, \citenamefont {Schleicher}, \citenamefont
  {Gruber}, \citenamefont {Avedissian}, \citenamefont {Arabski}, \citenamefont
  {Da~Costa}, \citenamefont {Schmerber}, \citenamefont {Rengasamy},
  \citenamefont {Vileno} \emph {et~al.}}]{boukari2018disentangling}%
  \BibitemOpen
  \bibfield  {author} {\bibinfo {author} {\bibfnamefont {S.}~\bibnamefont
  {Boukari}}, \bibinfo {author} {\bibfnamefont {H.}~\bibnamefont {Jabbar}},
  \bibinfo {author} {\bibfnamefont {F.}~\bibnamefont {Schleicher}}, \bibinfo
  {author} {\bibfnamefont {M.}~\bibnamefont {Gruber}}, \bibinfo {author}
  {\bibfnamefont {G.}~\bibnamefont {Avedissian}}, \bibinfo {author}
  {\bibfnamefont {J.}~\bibnamefont {Arabski}}, \bibinfo {author} {\bibfnamefont
  {V.}~\bibnamefont {Da~Costa}}, \bibinfo {author} {\bibfnamefont
  {G.}~\bibnamefont {Schmerber}}, \bibinfo {author} {\bibfnamefont
  {P.}~\bibnamefont {Rengasamy}}, \bibinfo {author} {\bibfnamefont
  {B.}~\bibnamefont {Vileno}}, \emph {et~al.},\ }\bibfield  {title} {\bibinfo
  {title} {Disentangling magnetic hardening and molecular spin chain
  contributions to exchange bias in ferromagnet/molecule bilayers},\
  }\href@noop {} {\bibfield  {journal} {\bibinfo  {journal} {Nano Letters}\
  }\textbf {\bibinfo {volume} {18}},\ \bibinfo {pages} {4659} (\bibinfo {year}
  {2018})}\BibitemShut {NoStop}%
\bibitem [{\citenamefont {Jo}\ \emph {et~al.}(2018)\citenamefont {Jo},
  \citenamefont {Byun}, \citenamefont {Oh}, \citenamefont {Park}, \citenamefont
  {Jin}, \citenamefont {Min}, \citenamefont {Lee},\ and\ \citenamefont
  {Yoo}}]{jo2018molecular}%
  \BibitemOpen
  \bibfield  {author} {\bibinfo {author} {\bibfnamefont {J.}~\bibnamefont
  {Jo}}, \bibinfo {author} {\bibfnamefont {J.}~\bibnamefont {Byun}}, \bibinfo
  {author} {\bibfnamefont {I.}~\bibnamefont {Oh}}, \bibinfo {author}
  {\bibfnamefont {J.}~\bibnamefont {Park}}, \bibinfo {author} {\bibfnamefont
  {M.-J.}\ \bibnamefont {Jin}}, \bibinfo {author} {\bibfnamefont {B.-C.}\
  \bibnamefont {Min}}, \bibinfo {author} {\bibfnamefont {J.}~\bibnamefont
  {Lee}},\ and\ \bibinfo {author} {\bibfnamefont {J.-W.}\ \bibnamefont {Yoo}},\
  }\bibfield  {title} {\bibinfo {title} {Molecular tunability of magnetic
  exchange bias and asymmetrical magnetotransport in
  metalloporphyrin/ {Co} hybrid bilayers},\ }\href@noop {}
  {\bibfield  {journal} {\bibinfo  {journal} {ACS nano}\ }\textbf {\bibinfo
  {volume} {13}},\ \bibinfo {pages} {894} (\bibinfo {year} {2018})}\BibitemShut
  {NoStop}%
\bibitem [{\citenamefont {Jo}\ \emph {et~al.}(2020)\citenamefont {Jo},
  \citenamefont {Byun}, \citenamefont {Lee}, \citenamefont {Choe},
  \citenamefont {Oh}, \citenamefont {Park}, \citenamefont {Jin}, \citenamefont
  {Lee},\ and\ \citenamefont {Yoo}}]{jo2020emergence}%
  \BibitemOpen
  \bibfield  {author} {\bibinfo {author} {\bibfnamefont {J.}~\bibnamefont
  {Jo}}, \bibinfo {author} {\bibfnamefont {J.}~\bibnamefont {Byun}}, \bibinfo
  {author} {\bibfnamefont {J.}~\bibnamefont {Lee}}, \bibinfo {author}
  {\bibfnamefont {D.}~\bibnamefont {Choe}}, \bibinfo {author} {\bibfnamefont
  {I.}~\bibnamefont {Oh}}, \bibinfo {author} {\bibfnamefont {J.}~\bibnamefont
  {Park}}, \bibinfo {author} {\bibfnamefont {M.-J.}\ \bibnamefont {Jin}},
  \bibinfo {author} {\bibfnamefont {J.}~\bibnamefont {Lee}},\ and\ \bibinfo
  {author} {\bibfnamefont {J.-W.}\ \bibnamefont {Yoo}},\ }\bibfield  {title}
  {\bibinfo {title} {Emergence of multispinterface and antiferromagnetic
  molecular exchange bias via molecular stacking on a ferromagnetic film},\
  }\href@noop {} {\bibfield  {journal} {\bibinfo  {journal} {Advanced
  Functional Materials}\ }\textbf {\bibinfo {volume} {30}},\ \bibinfo {pages}
  {1908499} (\bibinfo {year} {2020})}\BibitemShut {NoStop}%
\bibitem [{\citenamefont {Mundlia}\ \emph {et~al.}(2020)\citenamefont
  {Mundlia}, \citenamefont {Chaudhary}, \citenamefont {Peri}, \citenamefont
  {Bhardwaj}, \citenamefont {Panda}, \citenamefont {Sasmal},\ and\
  \citenamefont {Raman}}]{PhysRevApplied.14.024095}%
  \BibitemOpen
  \bibfield  {author} {\bibinfo {author} {\bibfnamefont {S.}~\bibnamefont
  {Mundlia}}, \bibinfo {author} {\bibfnamefont {S.}~\bibnamefont {Chaudhary}},
  \bibinfo {author} {\bibfnamefont {L.}~\bibnamefont {Peri}}, \bibinfo {author}
  {\bibfnamefont {A.}~\bibnamefont {Bhardwaj}}, \bibinfo {author}
  {\bibfnamefont {J.~J.}\ \bibnamefont {Panda}}, \bibinfo {author}
  {\bibfnamefont {S.}~\bibnamefont {Sasmal}},\ and\ \bibinfo {author}
  {\bibfnamefont {K.~V.}\ \bibnamefont {Raman}},\ }\bibfield  {title} {\bibinfo
  {title} {Robust monolayer exchange-bias effect in molecular crane-pulley
  response at magnetic surface},\ }\href
  {https://doi.org/10.1103/PhysRevApplied.14.024095} {\bibfield  {journal}
  {\bibinfo  {journal} {Phys. Rev. Applied}\ }\textbf {\bibinfo {volume}
  {14}},\ \bibinfo {pages} {024095} (\bibinfo {year} {2020})}\BibitemShut
  {NoStop}%
\bibitem [{\citenamefont {Sanvito}(2010)}]{sanvito2010molecular}%
  \BibitemOpen
  \bibfield  {author} {\bibinfo {author} {\bibfnamefont {S.}~\bibnamefont
  {Sanvito}},\ }\bibfield  {title} {\bibinfo {title} {Molecular spintronics:
  The rise of spinterface science},\ }\href@noop {} {\bibfield  {journal}
  {\bibinfo  {journal} {Nature Physics}\ }\textbf {\bibinfo {volume} {6}},\
  \bibinfo {pages} {562} (\bibinfo {year} {2010})}\BibitemShut {NoStop}%
\bibitem [{\citenamefont {Raman}(2014)}]{raman2014interface}%
  \BibitemOpen
  \bibfield  {author} {\bibinfo {author} {\bibfnamefont {K.~V.}\ \bibnamefont
  {Raman}},\ }\bibfield  {title} {\bibinfo {title} {Interface-assisted
  molecular spintronics},\ }\href@noop {} {\bibfield  {journal} {\bibinfo
  {journal} {Applied Physics Reviews}\ }\textbf {\bibinfo {volume} {1}},\
  \bibinfo {pages} {031101} (\bibinfo {year} {2014})}\BibitemShut {NoStop}%
\bibitem [{\citenamefont {Atodiresei}\ and\ \citenamefont
  {Raman}(2014)}]{atodiresei2014interface}%
  \BibitemOpen
  \bibfield  {author} {\bibinfo {author} {\bibfnamefont {N.}~\bibnamefont
  {Atodiresei}}\ and\ \bibinfo {author} {\bibfnamefont {K.~V.}\ \bibnamefont
  {Raman}},\ }\bibfield  {title} {\bibinfo {title} {Interface-assisted
  spintronics: Tailoring at the molecular scale},\ }\href@noop {} {\bibfield
  {journal} {\bibinfo  {journal} {MRS Bull}\ }\textbf {\bibinfo {volume}
  {39}},\ \bibinfo {pages} {596} (\bibinfo {year} {2014})}\BibitemShut
  {NoStop}%
\bibitem [{\citenamefont {Atodiresei}\ \emph {et~al.}(2013)\citenamefont
  {Atodiresei}, \citenamefont {Caciuc}, \citenamefont {Lazic},\ and\
  \citenamefont {Bl{\"u}gel}}]{Atodiresei2013ChemicalAV}%
  \BibitemOpen
  \bibfield  {author} {\bibinfo {author} {\bibfnamefont {N.}~\bibnamefont
  {Atodiresei}}, \bibinfo {author} {\bibfnamefont {V.}~\bibnamefont {Caciuc}},
  \bibinfo {author} {\bibfnamefont {P.}~\bibnamefont {Lazic}},\ and\ \bibinfo
  {author} {\bibfnamefont {S.}~\bibnamefont {Bl{\"u}gel}},\ }\bibfield  {title}
  {\bibinfo {title} {Chemical and van der waals interactions at hybrid
  organic-metal interfaces}\ }(\bibinfo {year} {2013})\BibitemShut {NoStop}%
\bibitem [{\citenamefont {Callsen}\ \emph {et~al.}(2013)\citenamefont
  {Callsen}, \citenamefont {Caciuc}, \citenamefont {Kiselev}, \citenamefont
  {Atodiresei},\ and\ \citenamefont {Bl\"ugel}}]{PhysRevLett.111.106805}%
  \BibitemOpen
  \bibfield  {author} {\bibinfo {author} {\bibfnamefont {M.}~\bibnamefont
  {Callsen}}, \bibinfo {author} {\bibfnamefont {V.}~\bibnamefont {Caciuc}},
  \bibinfo {author} {\bibfnamefont {N.}~\bibnamefont {Kiselev}}, \bibinfo
  {author} {\bibfnamefont {N.}~\bibnamefont {Atodiresei}},\ and\ \bibinfo
  {author} {\bibfnamefont {S.}~\bibnamefont {Bl\"ugel}},\ }\bibfield  {title}
  {\bibinfo {title} {Magnetic hardening induced by nonmagnetic organic
  molecules},\ }\href {https://doi.org/10.1103/PhysRevLett.111.106805}
  {\bibfield  {journal} {\bibinfo  {journal} {Phys. Rev. Lett.}\ }\textbf
  {\bibinfo {volume} {111}},\ \bibinfo {pages} {106805} (\bibinfo {year}
  {2013})}\BibitemShut {NoStop}%
\bibitem [{\citenamefont {Raman}\ \emph {et~al.}(2013)\citenamefont {Raman},
  \citenamefont {Kamerbeek}, \citenamefont {Mukherjee}, \citenamefont
  {Atodiresei}, \citenamefont {Sen}, \citenamefont {Lazi{\'c}}, \citenamefont
  {Caciuc}, \citenamefont {Michel}, \citenamefont {Stalke}, \citenamefont
  {Mandal} \emph {et~al.}}]{raman2013interface}%
  \BibitemOpen
  \bibfield  {author} {\bibinfo {author} {\bibfnamefont {K.~V.}\ \bibnamefont
  {Raman}}, \bibinfo {author} {\bibfnamefont {A.~M.}\ \bibnamefont
  {Kamerbeek}}, \bibinfo {author} {\bibfnamefont {A.}~\bibnamefont
  {Mukherjee}}, \bibinfo {author} {\bibfnamefont {N.}~\bibnamefont
  {Atodiresei}}, \bibinfo {author} {\bibfnamefont {T.~K.}\ \bibnamefont {Sen}},
  \bibinfo {author} {\bibfnamefont {P.}~\bibnamefont {Lazi{\'c}}}, \bibinfo
  {author} {\bibfnamefont {V.}~\bibnamefont {Caciuc}}, \bibinfo {author}
  {\bibfnamefont {R.}~\bibnamefont {Michel}}, \bibinfo {author} {\bibfnamefont
  {D.}~\bibnamefont {Stalke}}, \bibinfo {author} {\bibfnamefont {S.~K.}\
  \bibnamefont {Mandal}}, \emph {et~al.},\ }\bibfield  {title} {\bibinfo
  {title} {Interface-engineered templates for molecular spin memory devices},\
  }\href@noop {} {\bibfield  {journal} {\bibinfo  {journal} {Nature}\ }\textbf
  {\bibinfo {volume} {493}},\ \bibinfo {pages} {509} (\bibinfo {year}
  {2013})}\BibitemShut {NoStop}%
\bibitem [{\citenamefont {Decker}\ \emph {et~al.}(2013)\citenamefont {Decker},
  \citenamefont {Brede}, \citenamefont {Atodiresei}, \citenamefont {Caciuc},
  \citenamefont {Bl\"ugel},\ and\ \citenamefont
  {Wiesendanger}}]{PhysRevB.87.041403}%
  \BibitemOpen
  \bibfield  {author} {\bibinfo {author} {\bibfnamefont {R.}~\bibnamefont
  {Decker}}, \bibinfo {author} {\bibfnamefont {J.}~\bibnamefont {Brede}},
  \bibinfo {author} {\bibfnamefont {N.}~\bibnamefont {Atodiresei}}, \bibinfo
  {author} {\bibfnamefont {V.}~\bibnamefont {Caciuc}}, \bibinfo {author}
  {\bibfnamefont {S.}~\bibnamefont {Bl\"ugel}},\ and\ \bibinfo {author}
  {\bibfnamefont {R.}~\bibnamefont {Wiesendanger}},\ }\bibfield  {title}
  {\bibinfo {title} {Atomic-scale magnetism of cobalt-intercalated graphene},\
  }\href {https://doi.org/10.1103/PhysRevB.87.041403} {\bibfield  {journal}
  {\bibinfo  {journal} {Phys. Rev. B}\ }\textbf {\bibinfo {volume} {87}},\
  \bibinfo {pages} {041403} (\bibinfo {year} {2013})}\BibitemShut {NoStop}%
\bibitem [{\citenamefont {Bairagi}\ \emph {et~al.}(2015)\citenamefont
  {Bairagi}, \citenamefont {Bellec}, \citenamefont {Repain}, \citenamefont
  {Chacon}, \citenamefont {Girard}, \citenamefont {Garreau}, \citenamefont
  {Lagoute}, \citenamefont {Rousset}, \citenamefont {Breitwieser},
  \citenamefont {Hu}, \citenamefont {Chao}, \citenamefont {Pai}, \citenamefont
  {Li}, \citenamefont {Smogunov},\ and\ \citenamefont
  {Barreteau}}]{PhysRevLett.114.247203}%
  \BibitemOpen
  \bibfield  {author} {\bibinfo {author} {\bibfnamefont {K.}~\bibnamefont
  {Bairagi}}, \bibinfo {author} {\bibfnamefont {A.}~\bibnamefont {Bellec}},
  \bibinfo {author} {\bibfnamefont {V.}~\bibnamefont {Repain}}, \bibinfo
  {author} {\bibfnamefont {C.}~\bibnamefont {Chacon}}, \bibinfo {author}
  {\bibfnamefont {Y.}~\bibnamefont {Girard}}, \bibinfo {author} {\bibfnamefont
  {Y.}~\bibnamefont {Garreau}}, \bibinfo {author} {\bibfnamefont
  {J.}~\bibnamefont {Lagoute}}, \bibinfo {author} {\bibfnamefont
  {S.}~\bibnamefont {Rousset}}, \bibinfo {author} {\bibfnamefont
  {R.}~\bibnamefont {Breitwieser}}, \bibinfo {author} {\bibfnamefont {Y.-C.}\
  \bibnamefont {Hu}}, \bibinfo {author} {\bibfnamefont {Y.~C.}\ \bibnamefont
  {Chao}}, \bibinfo {author} {\bibfnamefont {W.~W.}\ \bibnamefont {Pai}},
  \bibinfo {author} {\bibfnamefont {D.}~\bibnamefont {Li}}, \bibinfo {author}
  {\bibfnamefont {A.}~\bibnamefont {Smogunov}},\ and\ \bibinfo {author}
  {\bibfnamefont {C.}~\bibnamefont {Barreteau}},\ }\bibfield  {title} {\bibinfo
  {title} {Tuning the magnetic anisotropy at a molecule-metal interface},\
  }\href {https://doi.org/10.1103/PhysRevLett.114.247203} {\bibfield  {journal}
  {\bibinfo  {journal} {Phys. Rev. Lett.}\ }\textbf {\bibinfo {volume} {114}},\
  \bibinfo {pages} {247203} (\bibinfo {year} {2015})}\BibitemShut {NoStop}%
\bibitem [{\citenamefont {Moorsom}\ \emph {et~al.}(2020)\citenamefont
  {Moorsom}, \citenamefont {Alghamdi}, \citenamefont {Stansill}, \citenamefont
  {Poli}, \citenamefont {Teobaldi}, \citenamefont {Beg}, \citenamefont
  {Fangohr}, \citenamefont {Rogers}, \citenamefont {Aslam}, \citenamefont
  {Ali}, \citenamefont {Hickey},\ and\ \citenamefont
  {Cespedes}}]{PhysRevB.101.060408}%
  \BibitemOpen
  \bibfield  {author} {\bibinfo {author} {\bibfnamefont {T.}~\bibnamefont
  {Moorsom}}, \bibinfo {author} {\bibfnamefont {S.}~\bibnamefont {Alghamdi}},
  \bibinfo {author} {\bibfnamefont {S.}~\bibnamefont {Stansill}}, \bibinfo
  {author} {\bibfnamefont {E.}~\bibnamefont {Poli}}, \bibinfo {author}
  {\bibfnamefont {G.}~\bibnamefont {Teobaldi}}, \bibinfo {author}
  {\bibfnamefont {M.}~\bibnamefont {Beg}}, \bibinfo {author} {\bibfnamefont
  {H.}~\bibnamefont {Fangohr}}, \bibinfo {author} {\bibfnamefont
  {M.}~\bibnamefont {Rogers}}, \bibinfo {author} {\bibfnamefont
  {Z.}~\bibnamefont {Aslam}}, \bibinfo {author} {\bibfnamefont
  {M.}~\bibnamefont {Ali}}, \bibinfo {author} {\bibfnamefont {B.~J.}\
  \bibnamefont {Hickey}},\ and\ \bibinfo {author} {\bibfnamefont
  {O.}~\bibnamefont {Cespedes}},\ }\bibfield  {title} {\bibinfo {title}
  {$\ensuremath{\pi}$-anisotropy: A nanocarbon route to hard magnetism},\
  }\href {https://doi.org/10.1103/PhysRevB.101.060408} {\bibfield  {journal}
  {\bibinfo  {journal} {Phys. Rev. B}\ }\textbf {\bibinfo {volume} {101}},\
  \bibinfo {pages} {060408} (\bibinfo {year} {2020})}\BibitemShut {NoStop}%
\bibitem [{\citenamefont {Nogu{\'e}s}\ and\ \citenamefont
  {Schuller}(1999)}]{nogues1999exchange}%
  \BibitemOpen
  \bibfield  {author} {\bibinfo {author} {\bibfnamefont {J.}~\bibnamefont
  {Nogu{\'e}s}}\ and\ \bibinfo {author} {\bibfnamefont {I.~K.}\ \bibnamefont
  {Schuller}},\ }\bibfield  {title} {\bibinfo {title} {Exchange bias},\
  }\href@noop {} {\bibfield  {journal} {\bibinfo  {journal} {Journal of
  Magnetism and Magnetic Materials}\ }\textbf {\bibinfo {volume} {192}},\
  \bibinfo {pages} {203} (\bibinfo {year} {1999})}\BibitemShut {NoStop}%
\bibitem [{\citenamefont {Zhang}\ and\ \citenamefont
  {Krishnan}(2016)}]{zhang2016epitaxial}%
  \BibitemOpen
  \bibfield  {author} {\bibinfo {author} {\bibfnamefont {W.}~\bibnamefont
  {Zhang}}\ and\ \bibinfo {author} {\bibfnamefont {K.~M.}\ \bibnamefont
  {Krishnan}},\ }\bibfield  {title} {\bibinfo {title} {Epitaxial exchange-bias
  systems: From fundamentals to future spin-orbitronics},\ }\href@noop {}
  {\bibfield  {journal} {\bibinfo  {journal} {Materials Science and
  Engineering: R: Reports}\ }\textbf {\bibinfo {volume} {105}},\ \bibinfo
  {pages} {1} (\bibinfo {year} {2016})}\BibitemShut {NoStop}%
\bibitem [{\citenamefont {Meiklejohn}\ and\ \citenamefont
  {Bean}(1957)}]{PhysRev.105.904}%
  \BibitemOpen
  \bibfield  {author} {\bibinfo {author} {\bibfnamefont {W.~H.}\ \bibnamefont
  {Meiklejohn}}\ and\ \bibinfo {author} {\bibfnamefont {C.~P.}\ \bibnamefont
  {Bean}},\ }\bibfield  {title} {\bibinfo {title} {New magnetic anisotropy},\
  }\href {https://doi.org/10.1103/PhysRev.105.904} {\bibfield  {journal}
  {\bibinfo  {journal} {Phys. Rev.}\ }\textbf {\bibinfo {volume} {105}},\
  \bibinfo {pages} {904} (\bibinfo {year} {1957})}\BibitemShut {NoStop}%
\bibitem [{\citenamefont {Yuan}\ \emph {et~al.}(2016)\citenamefont {Yuan},
  \citenamefont {Su}, \citenamefont {Song}, \citenamefont {Xing}, \citenamefont
  {Chen}, \citenamefont {Wang}, \citenamefont {Zhang}, \citenamefont {Ma},
  \citenamefont {Gao}, \citenamefont {Shi} \emph {et~al.}}]{yuan2016crystal}%
  \BibitemOpen
  \bibfield  {author} {\bibinfo {author} {\bibfnamefont {W.}~\bibnamefont
  {Yuan}}, \bibinfo {author} {\bibfnamefont {T.}~\bibnamefont {Su}}, \bibinfo
  {author} {\bibfnamefont {Q.}~\bibnamefont {Song}}, \bibinfo {author}
  {\bibfnamefont {W.}~\bibnamefont {Xing}}, \bibinfo {author} {\bibfnamefont
  {Y.}~\bibnamefont {Chen}}, \bibinfo {author} {\bibfnamefont {T.}~\bibnamefont
  {Wang}}, \bibinfo {author} {\bibfnamefont {Z.}~\bibnamefont {Zhang}},
  \bibinfo {author} {\bibfnamefont {X.}~\bibnamefont {Ma}}, \bibinfo {author}
  {\bibfnamefont {P.}~\bibnamefont {Gao}}, \bibinfo {author} {\bibfnamefont
  {J.}~\bibnamefont {Shi}}, \emph {et~al.},\ }\bibfield  {title} {\bibinfo
  {title} {Crystal structure manipulation of the exchange bias in an
  antiferromagnetic film},\ }\href@noop {} {\bibfield  {journal} {\bibinfo
  {journal} {Scientific reports}\ }\textbf {\bibinfo {volume} {6}},\ \bibinfo
  {pages} {28397} (\bibinfo {year} {2016})}\BibitemShut {NoStop}%
\bibitem [{\citenamefont {Ali}\ \emph {et~al.}(2003)\citenamefont {Ali},
  \citenamefont {Marrows}, \citenamefont {Al-Jawad}, \citenamefont {Hickey},
  \citenamefont {Misra}, \citenamefont {Nowak},\ and\ \citenamefont
  {Usadel}}]{PhysRevB.68.214420}%
  \BibitemOpen
  \bibfield  {author} {\bibinfo {author} {\bibfnamefont {M.}~\bibnamefont
  {Ali}}, \bibinfo {author} {\bibfnamefont {C.~H.}\ \bibnamefont {Marrows}},
  \bibinfo {author} {\bibfnamefont {M.}~\bibnamefont {Al-Jawad}}, \bibinfo
  {author} {\bibfnamefont {B.~J.}\ \bibnamefont {Hickey}}, \bibinfo {author}
  {\bibfnamefont {A.}~\bibnamefont {Misra}}, \bibinfo {author} {\bibfnamefont
  {U.}~\bibnamefont {Nowak}},\ and\ \bibinfo {author} {\bibfnamefont {K.~D.}\
  \bibnamefont {Usadel}},\ }\bibfield  {title} {\bibinfo {title}
  {Antiferromagnetic layer thickness dependence of the {IrMn/Co}
  exchange-bias system},\ }
  {\bibfield  {journal} {\bibinfo  {journal} {Phys. Rev. B}\ }\textbf {\bibinfo
  {volume} {68}},\ \bibinfo {pages} {214420} (\bibinfo {year}
  {2003})}\BibitemShut {NoStop}%
\bibitem [{\citenamefont {Nowak}\ \emph {et~al.}(2001)\citenamefont {Nowak},
  \citenamefont {Misra},\ and\ \citenamefont {Usadel}}]{nowak2001domain}%
  \BibitemOpen
  \bibfield  {author} {\bibinfo {author} {\bibfnamefont {U.}~\bibnamefont
  {Nowak}}, \bibinfo {author} {\bibfnamefont {A.}~\bibnamefont {Misra}},\ and\
  \bibinfo {author} {\bibfnamefont {K.~D.}\ \bibnamefont {Usadel}},\ }\bibfield
   {title} {\bibinfo {title} {Domain state model for exchange bias},\
  }\href@noop {} {\bibfield  {journal} {\bibinfo  {journal} {Journal of Applied
  Physics}\ }\textbf {\bibinfo {volume} {89}},\ \bibinfo {pages} {7269}
  (\bibinfo {year} {2001})}\BibitemShut {NoStop}%
\bibitem [{\citenamefont {Stiles}\ and\ \citenamefont
  {McMichael}(1999)}]{PhysRevB.59.3722}%
  \BibitemOpen
  \bibfield  {author} {\bibinfo {author} {\bibfnamefont {M.~D.}\ \bibnamefont
  {Stiles}}\ and\ \bibinfo {author} {\bibfnamefont {R.~D.}\ \bibnamefont
  {McMichael}},\ }\bibfield  {title} {\bibinfo {title} {Model for exchange bias
  in polycrystalline ferromagnet-antiferromagnet bilayers},\ }\href
  {https://doi.org/10.1103/PhysRevB.59.3722} {\bibfield  {journal} {\bibinfo
  {journal} {Phys. Rev. B}\ }\textbf {\bibinfo {volume} {59}},\ \bibinfo
  {pages} {3722} (\bibinfo {year} {1999})}\BibitemShut {NoStop}%
\bibitem [{\citenamefont {Moritz}\ \emph {et~al.}(2016)\citenamefont {Moritz},
  \citenamefont {Bacher},\ and\ \citenamefont {Dieny}}]{PhysRevB.94.104425}%
  \BibitemOpen
  \bibfield  {author} {\bibinfo {author} {\bibfnamefont {J.}~\bibnamefont
  {Moritz}}, \bibinfo {author} {\bibfnamefont {P.}~\bibnamefont {Bacher}},\
  and\ \bibinfo {author} {\bibfnamefont {B.}~\bibnamefont {Dieny}},\ }\bibfield
   {title} {\bibinfo {title} {Numerical study of the influence of interfacial
  roughness on the exchange bias properties of ferromagnetic/antiferromagnetic
  bilayers},\ }\href {https://doi.org/10.1103/PhysRevB.94.104425} {\bibfield
  {journal} {\bibinfo  {journal} {Phys. Rev. B}\ }\textbf {\bibinfo {volume}
  {94}},\ \bibinfo {pages} {104425} (\bibinfo {year} {2016})}\BibitemShut
  {NoStop}%
\bibitem [{MAI(2012)}]{MAITRE2012403}%
  \BibitemOpen
  \bibfield  {title} {\bibinfo {title} {Interfacial roughness and temperature
  effects on exchange bias properties in coupled
  ferromagnetic/antiferromagnetic bilayers},\ }\href@noop {} {\bibfield
  {journal} {\bibinfo  {journal} {Journal of Magnetism and Magnetic Materials}\
  }\textbf {\bibinfo {volume} {324}},\ \bibinfo {pages} {403 } (\bibinfo {year}
  {2012})}\BibitemShut {NoStop}%
\bibitem [{\citenamefont {Malozemoff}(1987)}]{PhysRevB.35.3679}%
  \BibitemOpen
  \bibfield  {author} {\bibinfo {author} {\bibfnamefont {A.~P.}\ \bibnamefont
  {Malozemoff}},\ }\bibfield  {title} {\bibinfo {title} {Random-field model of
  exchange anisotropy at rough ferromagnetic-antiferromagnetic interfaces},\
  }\href {https://doi.org/10.1103/PhysRevB.35.3679} {\bibfield  {journal}
  {\bibinfo  {journal} {Phys. Rev. B}\ }\textbf {\bibinfo {volume} {35}},\
  \bibinfo {pages} {3679} (\bibinfo {year} {1987})}\BibitemShut {NoStop}%
\bibitem [{\citenamefont {G\"okemeijer}\ \emph {et~al.}(2001)\citenamefont
  {G\"okemeijer}, \citenamefont {Penn}, \citenamefont {Veblen},\ and\
  \citenamefont {Chien}}]{PhysRevB.63.174422}%
  \BibitemOpen
  \bibfield  {author} {\bibinfo {author} {\bibfnamefont {N.~J.}\ \bibnamefont
  {G\"okemeijer}}, \bibinfo {author} {\bibfnamefont {R.~L.}\ \bibnamefont
  {Penn}}, \bibinfo {author} {\bibfnamefont {D.~R.}\ \bibnamefont {Veblen}},\
  and\ \bibinfo {author} {\bibfnamefont {C.~L.}\ \bibnamefont {Chien}},\
  }\bibfield  {title} {\bibinfo {title} {Exchange coupling in epitaxial
  {CoO/NiFe} bilayers with compensated and uncompensated
  interfacial spin structures},\ }\href
  {https://doi.org/10.1103/PhysRevB.63.174422} {\bibfield  {journal} {\bibinfo
  {journal} {Phys. Rev. B}\ }\textbf {\bibinfo {volume} {63}},\ \bibinfo
  {pages} {174422} (\bibinfo {year} {2001})}\BibitemShut {NoStop}%
\bibitem [{\citenamefont {Takano}\ \emph {et~al.}(1997)\citenamefont {Takano},
  \citenamefont {Kodama}, \citenamefont {Berkowitz}, \citenamefont {Cao},\ and\
  \citenamefont {Thomas}}]{PhysRevLett.79.1130}%
  \BibitemOpen
  \bibfield  {author} {\bibinfo {author} {\bibfnamefont {K.}~\bibnamefont
  {Takano}}, \bibinfo {author} {\bibfnamefont {R.~H.}\ \bibnamefont {Kodama}},
  \bibinfo {author} {\bibfnamefont {A.~E.}\ \bibnamefont {Berkowitz}}, \bibinfo
  {author} {\bibfnamefont {W.}~\bibnamefont {Cao}},\ and\ \bibinfo {author}
  {\bibfnamefont {G.}~\bibnamefont {Thomas}},\ }\bibfield  {title} {\bibinfo
  {title} {Interfacial uncompensated antiferromagnetic spins: Role in
  unidirectional anisotropy in polycrystalline
  {Ni$_{81}$Fe$_{19}$/CoO} bilayers},\ }\href
  {https://doi.org/10.1103/PhysRevLett.79.1130} {\bibfield  {journal} {\bibinfo
   {journal} {Phys. Rev. Lett.}\ }\textbf {\bibinfo {volume} {79}},\ \bibinfo
  {pages} {1130} (\bibinfo {year} {1997})}\BibitemShut {NoStop}%
\bibitem [{\citenamefont {Fan}\ \emph {et~al.}(2013)\citenamefont {Fan},
  \citenamefont {Smith}, \citenamefont {L{\"u}pke}, \citenamefont {Hanbicki},
  \citenamefont {Goswami}, \citenamefont {Li}, \citenamefont {Zhao},\ and\
  \citenamefont {Jonker}}]{fan2013exchange}%
  \BibitemOpen
  \bibfield  {author} {\bibinfo {author} {\bibfnamefont {Y.}~\bibnamefont
  {Fan}}, \bibinfo {author} {\bibfnamefont {K.}~\bibnamefont {Smith}}, \bibinfo
  {author} {\bibfnamefont {G.}~\bibnamefont {L{\"u}pke}}, \bibinfo {author}
  {\bibfnamefont {A.}~\bibnamefont {Hanbicki}}, \bibinfo {author}
  {\bibfnamefont {R.}~\bibnamefont {Goswami}}, \bibinfo {author} {\bibfnamefont
  {C.}~\bibnamefont {Li}}, \bibinfo {author} {\bibfnamefont {H.}~\bibnamefont
  {Zhao}},\ and\ \bibinfo {author} {\bibfnamefont {B.}~\bibnamefont {Jonker}},\
  }\bibfield  {title} {\bibinfo {title} {Exchange bias of the interface spin
  system at the {Fe/MgO} interface},\ }\href@noop {} {\bibfield
  {journal} {\bibinfo  {journal} {Nature nanotechnology}\ }\textbf {\bibinfo
  {volume} {8}},\ \bibinfo {pages} {438} (\bibinfo {year} {2013})}\BibitemShut
  {NoStop}%
\bibitem [{\citenamefont {Zhou}\ \emph {et~al.}(2017)\citenamefont {Zhou},
  \citenamefont {Song}, \citenamefont {Bai}, \citenamefont {Quan},
  \citenamefont {Jiang}, \citenamefont {Liu}, \citenamefont {Xu}, \citenamefont
  {Dhesi},\ and\ \citenamefont {Xu}}]{zhou2017robust}%
  \BibitemOpen
  \bibfield  {author} {\bibinfo {author} {\bibfnamefont {G.}~\bibnamefont
  {Zhou}}, \bibinfo {author} {\bibfnamefont {C.}~\bibnamefont {Song}}, \bibinfo
  {author} {\bibfnamefont {Y.}~\bibnamefont {Bai}}, \bibinfo {author}
  {\bibfnamefont {Z.}~\bibnamefont {Quan}}, \bibinfo {author} {\bibfnamefont
  {F.}~\bibnamefont {Jiang}}, \bibinfo {author} {\bibfnamefont
  {W.}~\bibnamefont {Liu}}, \bibinfo {author} {\bibfnamefont {Y.}~\bibnamefont
  {Xu}}, \bibinfo {author} {\bibfnamefont {S.~S.}\ \bibnamefont {Dhesi}},\ and\
  \bibinfo {author} {\bibfnamefont {X.}~\bibnamefont {Xu}},\ }\bibfield
  {title} {\bibinfo {title} {Robust interfacial exchange bias and
  metal--insulator transition influenced by the {LaNiO$_3$} layer
  thickness in {La$_{0.7}$Sr$_{0. 3}$MnO$_3$/LaNiO$_3$}
  superlattices},\ }\href@noop {} {\bibfield  {journal} {\bibinfo  {journal}
  {ACS applied materials \& interfaces}\ }\textbf {\bibinfo {volume} {9}},\
  \bibinfo {pages} {3156} (\bibinfo {year} {2017})}\BibitemShut {NoStop}%
\bibitem [{\citenamefont {Ali}\ \emph {et~al.}(2007)\citenamefont {Ali},
  \citenamefont {Adie}, \citenamefont {Marrows}, \citenamefont {Greig},
  \citenamefont {Hickey},\ and\ \citenamefont {Stamps}}]{ali2007exchange}%
  \BibitemOpen
  \bibfield  {author} {\bibinfo {author} {\bibfnamefont {M.}~\bibnamefont
  {Ali}}, \bibinfo {author} {\bibfnamefont {P.}~\bibnamefont {Adie}}, \bibinfo
  {author} {\bibfnamefont {C.~H.}\ \bibnamefont {Marrows}}, \bibinfo {author}
  {\bibfnamefont {D.}~\bibnamefont {Greig}}, \bibinfo {author} {\bibfnamefont
  {B.~J.}\ \bibnamefont {Hickey}},\ and\ \bibinfo {author} {\bibfnamefont
  {R.~L.}\ \bibnamefont {Stamps}},\ }\bibfield  {title} {\bibinfo {title}
  {Exchange bias using a spin glass},\ }\href@noop {} {\bibfield  {journal}
  {\bibinfo  {journal} {Nature Materials}\ }\textbf {\bibinfo {volume} {6}},\
  \bibinfo {pages} {70} (\bibinfo {year} {2007})}\BibitemShut {NoStop}%
\bibitem [{\citenamefont {Mauri}\ \emph {et~al.}(1987)\citenamefont {Mauri},
  \citenamefont {Siegmann}, \citenamefont {Bagus},\ and\ \citenamefont
  {Kay}}]{mauri1987simple}%
  \BibitemOpen
  \bibfield  {author} {\bibinfo {author} {\bibfnamefont {D.}~\bibnamefont
  {Mauri}}, \bibinfo {author} {\bibfnamefont {H.}~\bibnamefont {Siegmann}},
  \bibinfo {author} {\bibfnamefont {P.}~\bibnamefont {Bagus}},\ and\ \bibinfo
  {author} {\bibfnamefont {E.}~\bibnamefont {Kay}},\ }\bibfield  {title}
  {\bibinfo {title} {Simple model for thin ferromagnetic films exchange coupled
  to an antiferromagnetic substrate},\ }\href@noop {} {\bibfield  {journal}
  {\bibinfo  {journal} {Journal of Applied Physics}\ }\textbf {\bibinfo
  {volume} {62}},\ \bibinfo {pages} {3047} (\bibinfo {year}
  {1987})}\BibitemShut {NoStop}%
\bibitem [{\citenamefont {Stoner}\ and\ \citenamefont
  {Wohlfarth}(1948)}]{stoner1948mechanism}%
  \BibitemOpen
  \bibfield  {author} {\bibinfo {author} {\bibfnamefont {E.~C.}\ \bibnamefont
  {Stoner}}\ and\ \bibinfo {author} {\bibfnamefont {E.}~\bibnamefont
  {Wohlfarth}},\ }\bibfield  {title} {\bibinfo {title} {A mechanism of magnetic
  hysteresis in heterogeneous alloys},\ }\href@noop {} {\bibfield  {journal}
  {\bibinfo  {journal} {Philosophical Transactions of the Royal Society of
  London. Series A, Mathematical and Physical Sciences}\ }\textbf {\bibinfo
  {volume} {240}},\ \bibinfo {pages} {599} (\bibinfo {year}
  {1948})}\BibitemShut {NoStop}%
\bibitem [{\citenamefont {Wee}\ \emph {et~al.}(2001)\citenamefont {Wee},
  \citenamefont {Stamps},\ and\ \citenamefont {Camley}}]{wee2001temperature}%
  \BibitemOpen
  \bibfield  {author} {\bibinfo {author} {\bibfnamefont {L.}~\bibnamefont
  {Wee}}, \bibinfo {author} {\bibfnamefont {R.~L.}\ \bibnamefont {Stamps}},\
  and\ \bibinfo {author} {\bibfnamefont {R.~E.}\ \bibnamefont {Camley}},\
  }\bibfield  {title} {\bibinfo {title} {Temperature dependence of domain-wall
  bias and coercivity},\ }\href@noop {} {\bibfield  {journal} {\bibinfo
  {journal} {Journal of Applied Physics}\ }\textbf {\bibinfo {volume} {89}},\
  \bibinfo {pages} {6913} (\bibinfo {year} {2001})}\BibitemShut {NoStop}%
\bibitem [{\citenamefont {Nowak}\ \emph {et~al.}(2002)\citenamefont {Nowak},
  \citenamefont {Usadel}, \citenamefont {Keller}, \citenamefont {Milt\'enyi},
  \citenamefont {Beschoten},\ and\ \citenamefont
  {G\"untherodt}}]{PhysRevB.66.014430}%
  \BibitemOpen
  \bibfield  {author} {\bibinfo {author} {\bibfnamefont {U.}~\bibnamefont
  {Nowak}}, \bibinfo {author} {\bibfnamefont {K.~D.}\ \bibnamefont {Usadel}},
  \bibinfo {author} {\bibfnamefont {J.}~\bibnamefont {Keller}}, \bibinfo
  {author} {\bibfnamefont {P.}~\bibnamefont {Milt\'enyi}}, \bibinfo {author}
  {\bibfnamefont {B.}~\bibnamefont {Beschoten}},\ and\ \bibinfo {author}
  {\bibfnamefont {G.}~\bibnamefont {G\"untherodt}},\ }\bibfield  {title}
  {\bibinfo {title} {Domain state model for exchange bias. i. theory},\ }\href
  {https://doi.org/10.1103/PhysRevB.66.014430} {\bibfield  {journal} {\bibinfo
  {journal} {Phys. Rev. B}\ }\textbf {\bibinfo {volume} {66}},\ \bibinfo
  {pages} {014430} (\bibinfo {year} {2002})}\BibitemShut {NoStop}%
\bibitem [{\citenamefont {Schlenker}\ \emph {et~al.}(1986)\citenamefont
  {Schlenker}, \citenamefont {Parkin}, \citenamefont {Scott},\ and\
  \citenamefont {Howard}}]{schlenker1986magnetic}%
  \BibitemOpen
  \bibfield  {author} {\bibinfo {author} {\bibfnamefont {C.}~\bibnamefont
  {Schlenker}}, \bibinfo {author} {\bibfnamefont {S.}~\bibnamefont {Parkin}},
  \bibinfo {author} {\bibfnamefont {J.}~\bibnamefont {Scott}},\ and\ \bibinfo
  {author} {\bibfnamefont {K.}~\bibnamefont {Howard}},\ }\bibfield  {title}
  {\bibinfo {title} {Magnetic disorder in the exchange bias bilayered
  {FeNi-FeMn} system},\ }\href@noop {} {\bibfield  {journal}
  {\bibinfo  {journal} {Journal of magnetism and magnetic materials}\ }\textbf
  {\bibinfo {volume} {54}},\ \bibinfo {pages} {801} (\bibinfo {year}
  {1986})}\BibitemShut {NoStop}%
\bibitem [{\citenamefont {Mydosh}(1993)}]{mydosh1993spin}%
  \BibitemOpen
  \bibfield  {author} {\bibinfo {author} {\bibfnamefont {J.~A.}\ \bibnamefont
  {Mydosh}},\ }\href@noop {} {\emph {\bibinfo {title} {Spin glasses: an
  experimental introduction}}}\ (\bibinfo  {publisher} {CRC Press},\ \bibinfo
  {year} {1993})\BibitemShut {NoStop}%
\bibitem [{\citenamefont {Lu}\ and\ \citenamefont
  {Kobayashi}(2016)}]{lu2016optically}%
  \BibitemOpen
  \bibfield  {author} {\bibinfo {author} {\bibfnamefont {H.}~\bibnamefont
  {Lu}}\ and\ \bibinfo {author} {\bibfnamefont {N.}~\bibnamefont {Kobayashi}},\
  }\bibfield  {title} {\bibinfo {title} {Optically active porphyrin and
  phthalocyanine systems},\ }\href@noop {} {\bibfield  {journal} {\bibinfo
  {journal} {Chemical Reviews}\ }\textbf {\bibinfo {volume} {116}},\ \bibinfo
  {pages} {6184} (\bibinfo {year} {2016})}\BibitemShut {NoStop}%
\bibitem [{\citenamefont {Nayak}\ \emph {et~al.}(2020)\citenamefont {Nayak},
  \citenamefont {Manna}, \citenamefont {Vijayabaskaran}, \citenamefont {Singh},
  \citenamefont {Chelvane},\ and\ \citenamefont {Bedanta}}]{nayak2020exchange}%
  \BibitemOpen
  \bibfield  {author} {\bibinfo {author} {\bibfnamefont {S.}~\bibnamefont
  {Nayak}}, \bibinfo {author} {\bibfnamefont {P.~K.}\ \bibnamefont {Manna}},
  \bibinfo {author} {\bibfnamefont {T.}~\bibnamefont {Vijayabaskaran}},
  \bibinfo {author} {\bibfnamefont {B.~B.}\ \bibnamefont {Singh}}, \bibinfo
  {author} {\bibfnamefont {J.~A.}\ \bibnamefont {Chelvane}},\ and\ \bibinfo
  {author} {\bibfnamefont {S.}~\bibnamefont {Bedanta}},\ }\bibfield  {title}
  {\bibinfo {title} {Exchange bias in {Fe/Ir$_{20}$Mn$_{80}$}
  bilayers: Role of spin-glass like interface and ‘bulk’antiferromagnet
  spins},\ }\href@noop {} {\bibfield  {journal} {\bibinfo  {journal} {Journal
  of Magnetism and Magnetic Materials}\ }\textbf {\bibinfo {volume} {499}},\
  \bibinfo {pages} {166267} (\bibinfo {year} {2020})}\BibitemShut {NoStop}%
\bibitem [{\citenamefont {Ding}\ \emph {et~al.}(2013)\citenamefont {Ding},
  \citenamefont {Lebedev}, \citenamefont {Turner}, \citenamefont {Tian},
  \citenamefont {Hu}, \citenamefont {Seo}, \citenamefont {Panagopoulos},
  \citenamefont {Prellier}, \citenamefont {Van~Tendeloo},\ and\ \citenamefont
  {Wu}}]{ding2013interfacial}%
  \BibitemOpen
  \bibfield  {author} {\bibinfo {author} {\bibfnamefont {J.}~\bibnamefont
  {Ding}}, \bibinfo {author} {\bibfnamefont {O.}~\bibnamefont {Lebedev}},
  \bibinfo {author} {\bibfnamefont {S.}~\bibnamefont {Turner}}, \bibinfo
  {author} {\bibfnamefont {Y.}~\bibnamefont {Tian}}, \bibinfo {author}
  {\bibfnamefont {W.}~\bibnamefont {Hu}}, \bibinfo {author} {\bibfnamefont
  {J.}~\bibnamefont {Seo}}, \bibinfo {author} {\bibfnamefont {C.}~\bibnamefont
  {Panagopoulos}}, \bibinfo {author} {\bibfnamefont {W.}~\bibnamefont
  {Prellier}}, \bibinfo {author} {\bibfnamefont {G.}~\bibnamefont
  {Van~Tendeloo}},\ and\ \bibinfo {author} {\bibfnamefont {T.}~\bibnamefont
  {Wu}},\ }\bibfield  {title} {\bibinfo {title} {Interfacial spin glass state
  and exchange bias in manganite bilayers with competing magnetic orders},\
  }\href@noop {} {\bibfield  {journal} {\bibinfo  {journal} {Physical Review
  B}\ }\textbf {\bibinfo {volume} {87}},\ \bibinfo {pages} {054428} (\bibinfo
  {year} {2013})}\BibitemShut {NoStop}%
\bibitem [{\citenamefont {Xi}\ and\ \citenamefont
  {White}(2000)}]{xiantiferro2000}%
  \BibitemOpen
  \bibfield  {author} {\bibinfo {author} {\bibfnamefont {H.}~\bibnamefont
  {Xi}}\ and\ \bibinfo {author} {\bibfnamefont {R.}~\bibnamefont {White}},\
  }\bibfield  {title} {\bibinfo {title} {Antiferromagnetic thickness dependence
  of exchange biasing},\ }\href@noop {} {\bibfield  {journal} {\bibinfo
  {journal} {Physical Review B}\ }\textbf {\bibinfo {volume} {61}},\ \bibinfo
  {pages} {80} (\bibinfo {year} {2000})}\BibitemShut {NoStop}%
\bibitem [{\citenamefont {Binder}\ and\ \citenamefont
  {Young}(1986)}]{RevModPhys.58.801}%
  \BibitemOpen
  \bibfield  {author} {\bibinfo {author} {\bibfnamefont {K.}~\bibnamefont
  {Binder}}\ and\ \bibinfo {author} {\bibfnamefont {A.~P.}\ \bibnamefont
  {Young}},\ }\bibfield  {title} {\bibinfo {title} {Spin glasses: Experimental
  facts, theoretical concepts, and open questions},\ }\href
  {https://doi.org/10.1103/RevModPhys.58.801} {\bibfield  {journal} {\bibinfo
  {journal} {Rev. Mod. Phys.}\ }\textbf {\bibinfo {volume} {58}},\ \bibinfo
  {pages} {801} (\bibinfo {year} {1986})}\BibitemShut {NoStop}%
\bibitem [{\citenamefont {Pradheesh}\ \emph {et~al.}(2012)\citenamefont
  {Pradheesh}, \citenamefont {Nair}, \citenamefont {Kumar}, \citenamefont
  {Lamsal}, \citenamefont {Nirmala}, \citenamefont {Santhosh}, \citenamefont
  {Yelon}, \citenamefont {Malik}, \citenamefont {Sankaranarayanan},\ and\
  \citenamefont {Sethupathi}}]{pradheesh2012}%
  \BibitemOpen
  \bibfield  {author} {\bibinfo {author} {\bibfnamefont {R.}~\bibnamefont
  {Pradheesh}}, \bibinfo {author} {\bibfnamefont {H.~S.}\ \bibnamefont {Nair}},
  \bibinfo {author} {\bibfnamefont {C.~M.~N.}\ \bibnamefont {Kumar}}, \bibinfo
  {author} {\bibfnamefont {J.}~\bibnamefont {Lamsal}}, \bibinfo {author}
  {\bibfnamefont {R.}~\bibnamefont {Nirmala}}, \bibinfo {author} {\bibfnamefont
  {P.~N.}\ \bibnamefont {Santhosh}}, \bibinfo {author} {\bibfnamefont {W.~B.}\
  \bibnamefont {Yelon}}, \bibinfo {author} {\bibfnamefont {S.~K.}\ \bibnamefont
  {Malik}}, \bibinfo {author} {\bibfnamefont {V.}~\bibnamefont
  {Sankaranarayanan}},\ and\ \bibinfo {author} {\bibfnamefont {K.}~\bibnamefont
  {Sethupathi}},\ }\bibfield  {title} {\bibinfo {title} {Observation of spin
  glass state in weakly ferromagnetic {Sr$_2$FeCoO$_6$} double
  perovskite},\ }\href@noop {} {\bibfield  {journal} {\bibinfo  {journal}
  {Journal of Applied Physics}\ }\textbf {\bibinfo {volume} {111}},\ \bibinfo
  {pages} {053905} (\bibinfo {year} {2012})}\BibitemShut {NoStop}%
\bibitem [{\citenamefont {Sato}\ \emph {et~al.}(2001)\citenamefont {Sato},
  \citenamefont {Ando}, \citenamefont {Ogawa}, \citenamefont {Morimoto},\ and\
  \citenamefont {Ito}}]{PhysRevB.64.184432}%
  \BibitemOpen
  \bibfield  {author} {\bibinfo {author} {\bibfnamefont {T.}~\bibnamefont
  {Sato}}, \bibinfo {author} {\bibfnamefont {T.}~\bibnamefont {Ando}}, \bibinfo
  {author} {\bibfnamefont {T.}~\bibnamefont {Ogawa}}, \bibinfo {author}
  {\bibfnamefont {S.}~\bibnamefont {Morimoto}},\ and\ \bibinfo {author}
  {\bibfnamefont {A.}~\bibnamefont {Ito}},\ }\bibfield  {title} {\bibinfo
  {title} {Spin freezing and the ferromagnetic and reentrant spin-glass phases
  in a reentrant ferromagnet},\ }\href
  {https://doi.org/10.1103/PhysRevB.64.184432} {\bibfield  {journal} {\bibinfo
  {journal} {Phys. Rev. B}\ }\textbf {\bibinfo {volume} {64}},\ \bibinfo
  {pages} {184432} (\bibinfo {year} {2001})}\BibitemShut {NoStop}%
\bibitem [{\citenamefont {Fitzsimmons}\ \emph {et~al.}(2000)\citenamefont
  {Fitzsimmons}, \citenamefont {Yashar}, \citenamefont {Leighton},
  \citenamefont {Schuller}, \citenamefont {Nogués}, \citenamefont {Majkrzak},\
  and\ \citenamefont {Dura}}]{fitzsimmons2000relation}%
  \BibitemOpen
  \bibfield  {author} {\bibinfo {author} {\bibfnamefont {M.~R.}\ \bibnamefont
  {Fitzsimmons}}, \bibinfo {author} {\bibfnamefont {P.}~\bibnamefont {Yashar}},
  \bibinfo {author} {\bibfnamefont {C.}~\bibnamefont {Leighton}}, \bibinfo
  {author} {\bibfnamefont {I.~K.}\ \bibnamefont {Schuller}}, \bibinfo {author}
  {\bibfnamefont {J.}~\bibnamefont {Nogués}}, \bibinfo {author} {\bibfnamefont
  {C.~F.}\ \bibnamefont {Majkrzak}},\ and\ \bibinfo {author} {\bibfnamefont
  {J.~A.}\ \bibnamefont {Dura}},\ }\bibfield  {title} {\bibinfo {title}
  {Asymmetric magnetization reversal in exchange-biased hysteresis loops},\
  }\href@noop {} {\bibfield  {journal} {\bibinfo  {journal} {Phys. Rev. Lett.}\
  }\textbf {\bibinfo {volume} {84}},\ \bibinfo {pages} {3986} (\bibinfo {year}
  {2000})}\BibitemShut {NoStop}%
\bibitem [{\citenamefont {Krivorotov}\ \emph {et~al.}(2002)\citenamefont
  {Krivorotov}, \citenamefont {Leighton}, \citenamefont {Nogu\'{e}s},
  \citenamefont {Schuller},\ and\ \citenamefont
  {Dahlberg}}]{krivorotov2002relation}%
  \BibitemOpen
  \bibfield  {author} {\bibinfo {author} {\bibfnamefont {I.}~\bibnamefont
  {Krivorotov}}, \bibinfo {author} {\bibfnamefont {C.}~\bibnamefont
  {Leighton}}, \bibinfo {author} {\bibfnamefont {J.}~\bibnamefont
  {Nogu\'{e}s}}, \bibinfo {author} {\bibfnamefont {I.~K.}\ \bibnamefont
  {Schuller}},\ and\ \bibinfo {author} {\bibfnamefont {E.~D.}\ \bibnamefont
  {Dahlberg}},\ }\bibfield  {title} {\bibinfo {title} {Relation between
  exchange anisotropy and magnetization reversal asymmetry in
  {Fe/MnF$_2$} bilayers},\ }\href@noop {} {\bibfield  {journal}
  {\bibinfo  {journal} {Phys. Rev. B}\ }\textbf {\bibinfo {volume} {66}},\
  \bibinfo {pages} {100402R} (\bibinfo {year} {2002})}\BibitemShut {NoStop}%
\end{thebibliography}
%

\onecolumngrid

\end{document}